\pdfoutput=1
\documentclass[aps,prl,reprint,twocolumn,showpacs,superscriptaddress,nofootinbib]{revtex4-1}  
\usepackage{tabularx}
\makeatletter
\def\hlinewd#1{%
\noalign{\ifnum0=`}\fi\hrule \@height #1 %
\futurelet\reserved@a\@xhline}
\makeatother
\usepackage{scrextend}
\usepackage{amsmath}
\usepackage{amssymb}
\usepackage{epsfig}
\usepackage{graphicx}
\usepackage{hyperref}
\usepackage{dcolumn}   % needed for some tables
\usepackage{slashed}
\usepackage{color}
\usepackage{rotating}
\usepackage[margin=0.9in,a4paper]{geometry}
\usepackage[table,xcdraw,dvipsnames]{xcolor}
\usepackage[utf8]{inputenc}
\usepackage{colortbl}
\usepackage[normalem]{ulem}
\usepackage{mathrsfs} 

\definecolor{nicered}{rgb}{0.7,0.1,0.1}
\definecolor{nicegreen}{rgb}{0.1,0.5,0.1}
\definecolor{red}{rgb}{1.0, 0, 0}

\newcommand{\G}{{\cal G}}
\renewcommand{\P}{{\cal P}}

\newcommand{\SU}{{\rm SU}}

\renewcommand{\[}{\left[}
\renewcommand{\]}{\right]}
\renewcommand{\(}{\left(}
\renewcommand{\)}{\right)}

\newcommand{\bdm}{\begin{displaymath}}
\newcommand{\edm}{\end{displaymath}}
\newcommand{\bea}{\begin{eqnarray}}
\newcommand{\eea}{\end{eqnarray}}

\renewcommand\P{\mathcal P}

\newcommand{\Ud}{U^\dagger}
\newcommand{\Chid}{\chi^\dagger}
\newcommand{\dmu}{\partial_\mu}
\newcommand{\dnu}{\partial_\nu}
\newcommand{\dmuu}{\partial^\mu}
\newcommand{\dnuu}{\partial^\nu}

\newcommand{\Dmu}{D_\mu}

\newcommand{\pim}{\pi_-}
\newcommand{\pip}{\pi_+}
\newcommand{\piz}{\pi_0}
\definecolor{nicered}{rgb}{0.7,0.1,0.1}
\definecolor{nicegreen}{rgb}{0.1,0.5,0.1}
\definecolor{red}{rgb}{1.0, 0, 0}
\definecolor{niceblue}{rgb}{0,0,0.8}
\definecolor{red}{rgb}{1.0, 0, 0}
\hypersetup{colorlinks,citecolor= nicegreen,linkcolor= nicered,urlcolor=nicered}

\def\eq#1{{Eq.~(\ref{#1})}}

\def\fig#1{{Fig.~\ref{#1}}}

\def\vev#1{\left\langle #1\right\rangle}

\def\Re{\mbox{Re}\,}
\def\Tr{\mbox{Tr}\,}

\def\diag{\mbox{diag}\,}

\def\gsim{\raise0.3ex\hbox{$\;>$\kern-0.75em\raise-1.1ex\hbox{$\sim\;$}}}
\def\lsim{\raise0.3ex\hbox{$\;<$\kern-0.75em\raise-1.1ex\hbox{$\sim\;$}}}
\def\mb[#1]{\mathbf{#1}}
\renewcommand{\bar}{\overline}

\definecolor{LightCyan}{rgb}{0.88,1,1}
\definecolor{piggypink}{rgb}{0.99, 0.87, 0.9}
\definecolor{applegreen}{rgb}{0.55, 0.71, 0.0}
\definecolor{darkpastelgreen}{rgb}{0.01, 0.75, 0.24}
\definecolor{green-yellow}{rgb}{0.68, 1.0, 0.18}

\newcommand{\beq}{\begin{equation}}
\newcommand{\eeq}{\end{equation}}
\newcommand{\beqa}{\begin{eqnarray}}
\newcommand{\eeqa}{\end{eqnarray}}

\newcommand{\Sec}[1]{ \medskip \noindent {\sl \bfseries #1}}

\begin{document}
% ----------------- preprint numbers ------------------
%\begin{frontmatter}

%%%%%%%%%%%%%%%%%%%%%% 
%\Red{{\bf Only in the arXiv version:}
%\begin{flushright}
%{\footnotesize}
%\end{flushright}

% ------------- Title and authors ---------------------

\title{Breakdown of chiral perturbation theory for the axion hot dark matter bound}

\author{Luca Di Luzio}
\email{luca.diluzio@desy.de}
\affiliation{\small \it 
%Deutsches Elektronen-Synchrotron 
DESY, Notkestra\ss e 85, 
D-22607 Hamburg, Germany}
\affiliation{\small \it 
Dipartimento di Fisica e Astronomia `G.~Galilei', Universit\`a di Padova, Italy}
\affiliation{\small \it 
INFN Sezione di Padova, Via Francesco Marzolo 8, 35131 Padova, Italy}
\author{Guido Martinelli}
\email{guido.martinelli@roma1.infn.it}
\affiliation{\small \it Physics Department and INFN Sezione di Roma La Sapienza, Piazzale Aldo Moro 5, 00185 Roma, Italy
}
\author{Gioacchino Piazza}
\email{gioacchino.piazza@ijclab.in2p3.fr}
\affiliation{\small \it IJCLab, P\^{o}le Th\'{e}orie (B\^{a}t. 210), CNRS/IN2P3 et Universit\'{e}  Paris-Saclay, 91405 Orsay, France}

% ------------------------------------------------------
\begin{abstract}
\noindent
We show that the commonly adopted hot dark matter (HDM) bound on the axion mass 
$m_a \lesssim$ 1 eV is not reliable, since it 
is obtained by extrapolating the chiral expansion 
in a region where the effective field theory breaks down.  
This is explicitly shown via the calculation of the axion-pion thermalization rate 
at the next-to-leading order 
in chiral perturbation theory. 
We 
finally 
advocate a strategy for a sound   
extraction of the axion HDM bound  
via lattice QCD techniques.

\end{abstract}

\maketitle

%%%%%%%%%%%%%%%%%%%%%% 
\Sec{Introduction.}  
The axion 
originally emerged
as a low-energy remnant 
of the Peccei Quinn solution to the strong CP 
problem \cite{Peccei:1977hh,Peccei:1977ur,Wilczek:1977pj,Weinberg:1977ma}, 
but it also unavoidably contributes to the energy density 
of the Universe. 
There are two qualitatively different 
populations of relic axions, a non-thermal one 
comprising cold dark matter (DM) \cite{Preskill:1982cy,Abbott:1982af,Dine:1982ah,Davis:1986xc}, 
and a thermal axion population \cite{Turner:1986tb} 
which, while still relativistic,
would behave as extra dark radiation. 
Such hot dark matter (HDM) component 
contributes to the effective number of extra 
relativistic degrees of freedom \cite{Kolb:1990vq}
$\Delta N_{\rm eff} \simeq 4/7 \( 43 / [4 g_S(T_D)] \)^{4/3}$,  
with $g_S(T_D)$ the number of entropy degrees of freedom 
at the axion decoupling temperature, $T_D$. 
The value of 
$\Delta N_{\rm eff}$ is constrained by 
cosmic microwave background (CMB) experiments, 
such as the Planck satellite \cite{Akrami:2018vks,Aghanim:2018eyx},   
while planned CMB Stage 4 (CMB-S4) 
experiments \cite{Abazajian:2016yjj} will 
provide an observable window on the axion interactions. 

There are several processes that can keep the axion 
in thermal equilibrium with the Standard Model (SM) thermal bath.
From the standpoint of the 
axion 
solution to the strong CP problem, 
an unavoidable process arises from the 
model-independent 
coupling to gluons, 
$\frac{\alpha_s}{8\pi} \frac{a}{f_a} G \tilde G$.\footnote{Other 
thermalization channels arise from 
model-dependent axion couplings to photons \cite{Turner:1986tb}, 
SM quarks \cite{Salvio:2013iaa,Baumann:2016wac,Ferreira:2018vjj,Arias-Aragon:2020shv}
and leptons \cite{DEramo:2018vss}.}   
For $T_D \gtrsim 1$ GeV thermal axion production proceeds 
via its scatterings with gluons in the 
quark-gluon plasma \cite{Masso:2002np,Graf:2010tv}, 
while for $T_D \lesssim 1$ GeV 
processes involving pions and nucleons  
must be considered \cite{Berezhiani:1992rk,Chang:1993gm,Hannestad:2005df}. 
The latter, have the advantage 
of occurring very late in the thermal history, 
so that it is unlikely that the corresponding population of thermal axions 
could be diluted by inflation. 
The transition between the two regimes depends on 
the strength of the axion interactions set by 
$f_a$ or, equivalently, by $m_a \simeq 5.7 \times (10^6 \ \text{GeV} / f_a)$ eV, 
and it encompasses the range $m_a \in [0.01,0.1]$ eV 
(with heavier axions leading to lower decoupling temperatures).  
Although the transition region cannot be precisely determined 
due to the complications of the quark-hadron 
phase transition, for heavier axions approaching the eV scale  
the main thermalization channel is  
$a\pi \leftrightarrow \pi\pi$ 
\cite{Chang:1993gm,Hannestad:2005df}, 
with $T_D \lesssim 200$ MeV.  
In this regime, 
scatterings off nucleons are subdominant 
because of the 
exponential suppression in their number density.  

The highest attainable axion mass 
from cosmological constraints 
on extra relativistic degrees of freedom, 
also known as HDM bound,  
translate into $m_a \lesssim$ 1 eV \cite{Zyla:2020zbs}. 
Based on a leading-order (LO) axion-pion chiral 
effective field theory (EFT) 
analysis of the axion-pion thermalization rate
\cite{Chang:1993gm,Hannestad:2005df}, 
the axion HDM bound has been reconsidered 
in Refs.~\cite{Melchiorri:2007cd,Hannestad:2008js,Hannestad:2010yi,Archidiacono:2013cha,Giusarma:2014zza,DiValentino:2015zta,DiValentino:2015wba,Archidiacono:2015mda,Giare:2020vzo}, 
also in correlation with relic neutrinos. 
The most recent update \cite{Giare:2020vzo}
quotes a 95$\%$ CL bound that 
ranges from 
$m_a \lesssim$ 0.2 eV to 1 eV, 
depending 
on the used data set and assumed cosmological model. 
Although the axion 
mass range 
relevant for the HDM bound
is in generic tension with astrophysical constraints, 
the latter can be tamed in several respects.\footnote{Tree-level axion couplings 
to electrons are absent in KSVZ 
models \cite{Kim:1979if,Shifman:1979if}, 
thus relaxing the constraints from Red Giants 
and White Dwarfs. 
The axion coupling to photons, 
constrained by Horizontal Branch stars evolution, 
can be accidentally suppressed in certain KSVZ-like 
models \cite{Kaplan:1985dv,DiLuzio:2016sbl,DiLuzio:2017pfr}. 
Finally, the SN1987A bound on the axion coupling to nucleons  
can be considered less robust 
both from the astrophysical and experimental point of view 
\cite{Raffelt:1990yz,Chang:2018rso,Carenza:2019pxu,Bar:2019ifz}.} 

It is the purpose of this Letter to revisit the axion HDM bound in 
the context of the next-to-LO (NLO) axion-pion chiral EFT. 
This is motivated by the simple observation that the 
mean energy of pions (axions) in a heat bath of $T \simeq 100$ MeV
is $\vev{E} \equiv \rho / n \simeq 350$ MeV ($270$ MeV), thus questioning the validity of the 
chiral expansion for the scattering process $a\pi \leftrightarrow \pi\pi$. 
The latter is expected to fail for $\sqrt{s} \sim \vev{E_\pi} + \vev{E_a} \gtrsim 500$ MeV, 
corresponding to temperatures well below that of QCD deconfinement, 
which was estimated to be $T_c = 154 \pm 9$ MeV in Ref.~\cite{Bazavov:2011nk},  
see also \cite{Aoki:2006br,Borsanyi:2010bp}. 

In this work, we provide for the first time the formulation of the full axion-pion Lagrangian 
at NLO, including also derivative axion couplings to the 
pionic current (previous NLO studies 
only considered 
non-derivative axion-pion interactions \cite{Spalinski:1988az,diCortona:2015ldu}), and paying special attention to the 
issue of the axion-pion mixing. 
Next, we perform a NLO calculation of the $a\pi \leftrightarrow \pi\pi$ 
thermalization rate 
(that can be cast as an expansion in $T/f_\pi$, with $f_\pi \simeq 92$ MeV)
and show that the NLO correction saturates 
half of the LO contribution for $T_\chi \simeq 62$ MeV. 
The latter can be considered as 
the maximal temperature above which 
the chiral description 
breaks down for the process under consideration.  
On the other hand, the region from $T_\chi$ up to $T_c$, 
where chiral perturbation theory cannot be applied, 
turns out to be crucial 
for the extraction of the HDM bound and for assessing 
the sensitivity of future CMB experiments. 

We conclude with 
a proposal for extracting the axion-pion thermalization rate 
via a direct Lattice QCD calculation, in analogy to the 
well-studied case of $\pi$-$\pi$ scattering. 

\Sec{Axion-pion scattering at LO.} 
The construction of the LO axion-pion Lagrangian was discussed 
long ago 
in Refs.~\cite{DiVecchia:1980yfw,Georgi:1986df}. 
We recall here its basic ingredients (see also \cite{Chang:1993gm,DiLuzio:2020wdo}), 
in view of the extension at NLO.
Defining the pion Goldstone matrix $U= e^{i\pi^A\sigma^A/ f_\pi}$, 
with $f_\pi \simeq 92$ MeV, 
$\pi^A$ and $\sigma^A$ ($A=1,2,3$) denoting 
respectively the real pion fields and 
the Pauli matrices, 
the LO axion-pion interactions stem from 
\beq 
\label{eq:axionpionLO}
\mathscr{L}^{\rm LO}_{a\text{-}\pi} = 
\frac{f_\pi^2}{4} {\rm Tr}\left[ U\Chid_a +\chi_a \Ud  \right] +
\frac{\dmuu a}{2 f_a} {\rm Tr}\left[c_q \sigma^A\right] J^{A}_\mu \, , 
\eeq
where $\chi_a = 2 B_0 M_a$, 
in terms of the quark condensate $B_0$ and the `axion-dressed' 
quark mass matrix $M_a = e^{i \frac{a}{2 f_a} Q_a } M_q e^{i \frac{a}{2 f_a} Q_a }$, 
with $M_q = \diag (m_u,m_d)$ 
and $\Tr Q_a = 1$. 
The latter condition ensures 
that the axion field is transferred from the 
operator $\frac{\alpha_s}{8\pi} \frac{a}{f_a} G \tilde G$ 
to the phase of the quark mass matrix, 
via the quark axial field 
redefinition $q \to \exp(i\gamma_5\frac{a}{2f_a}Q_a)q$.
In the following, we set $Q_a = M_q^{-1} / \Tr M_q^{-1}$, 
so that 
terms linear in $a$ 
(including $a$-$\pi^0$ mass mixing)
drop out from the first term in \eq{eq:axionpionLO}. 
Hence, in this basis, 
the only linear axion interaction is the derivative one 
with the conserved $\SU(2)_A$ pion current.  
The latter reads 
at LO  
\beq 
J^{A}_\mu|^{\rm LO} = \frac{i}{4} f_\pi^2 {\rm Tr} \[\sigma^A \(U \partial_\mu \Ud -\Ud \partial_\mu U \) \] \, , 
\eeq
while the derivative axion coupling in \eq{eq:axionpionLO} 
is
$\Tr \[ c_q \sigma^A \] = 
(\frac{m_u - m_d}{m_u + m_d} 
+ c^0_{u} - c^0_{d}) \delta^{A3}$, 
where the first term 
arises from the axial quark rotation that removed 
the $aG\tilde G$ operator 
and the second one 
originates from
the model-dependent coefficient 
$c^0_{q} = \text{diag}(c^0_{u},c^0_{d})$,  
defined via the Lagrangian term 
$\frac{\partial^\mu a}{2 f_a} \bar q c^0_{q} \gamma_\mu \gamma_5 q$. 
For instance, $c^0_{u,d} = 0$ in the KSVZ model \cite{Kim:1979if,Shifman:1979if}, 
while $c^0_{u} = \frac{1}{3} \cos^2\beta$ and $c^0_{d} = \frac{1}{3} \sin^2\beta$ 
in the DFSZ model \cite{Zhitnitsky:1980tq,Dine:1981rt}, 
with $\tan\beta$ the ratio between the 
vacuum expectation values 
of two Higgs doublets.
Expanding the pion matrix in \eq{eq:axionpionLO} one obtains 
\beq
\label{eq:axionpionLOexp}
\mathscr{L}^{\rm LO}_{a\text{-}\pi} \supset 
\epsilon \,
\partial^\mu a \partial_\mu \piz + \frac{C_{a\pi}}{f_a f_\pi} \partial^\mu a
[\partial \pi\pi\pi]_\mu 
\, ,
\eeq
with the definitions $[\partial \pi\pi\pi]_\mu = 2 \dmu \piz \pip \pim -\piz \dmu \pip \pim -\piz \pip \dmu\pim$, 
$\epsilon = - \frac{3 f_\pi C_{a\pi}}{2 f_a}$ and
\beq 
\label{eq:defCapi}
C_{a\pi} = \frac{1}{3}\( \frac{m_d-m_u}{m_u+m_d} +c_d^0-c_u^0\) \, .
\eeq
At the LO in $\epsilon$ the diagonalization of the $a$-$\pi^0$ 
term 
is obtained by shifting 
$a \to a - \epsilon \pi^0$ and $\pi^0 \to \pi^0 + \mathcal{O}(\epsilon^3) a$,  
where we used the fact that $m_a / m_\pi = \mathcal{O}(\epsilon)$. 
Hence, as long as we are interested in effects 
that are linear in $a$ 
and neglect $\mathcal{O}(\epsilon^3)$ 
corrections,  
the axion-pion 
interactions in \eq{eq:axionpionLOexp} 
are already in the basis with canonical propagators. 

For temperatures below the QCD phase transition, 
the main processes relevant for the 
axion thermalization rate are 
$a (p_1) \piz (p_2)\rightarrow\pip (p_3) \pim (p_4)$, 
whose amplitude at LO reads 
\beq 
\label{eq:Mapi0pippimLO} 
\mathcal{M}^{\rm LO}_{a \piz \rightarrow \pip \pim} =
\frac{C_{a\pi}}{f_\pi f_a} \frac{3}{2}\left[m_\pi^2-s\right] \, , 
\eeq
with $s= (p_1 + p_2)^2$, 
together with the crossed channels 
$a \pim\rightarrow\piz \pim$ and $a \pip\rightarrow\pip \piz$.   
The amplitudes of the latter are obtained by replacing $s \to t=(p_1 - p_3)^2$ 
and $s \to u=(p_1 - p_4)^2$, respectively. 
Taking equal masses for the neutral and charged pions, 
one finds the squared matrix element (summed over the three channels above) \cite{Hannestad:2005df} 
\begin{equation}
\sum|\mathcal{M}|_{\rm LO}^2= \left(\frac{C_{a\pi}}{f_a f_\pi}\right)^2\frac{9}{4}\left[s^2+t^2+u^2-3m_\pi^4\right] \, . 
\end{equation}

\Sec{Axion-pion scattering at NLO.} 
To compute the axion thermalization process beyond LO we need to consider the one-loop amplitudes from the LO Lagrangian in \eq{eq:axionpionLO} as well as the tree-level amplitudes stemming from the NLO axion-pion Lagrangian, both contributing to $\mathcal{O}(p^4)$ in the chiral expansion. 
The NLO interactions include the derivative coupling of the axion to the NLO axial current, 
which has been computed here for the first time.  

We stick to the expression of the NLO chiral Lagrangian given in Ref.~\cite{Gasser:1983yg} 
(see for example Appendix D in \cite{Scherer:2002tk} for trace notation), 
which, considering only two flavours, 
depends on $10$ low-energy constants (LECs) 
$\ell_1,\ell_2, \dots, \ell_7,h_1,h_2,h_3$. 
The axion field has been included in the phase of the quark mass matrix, as described after \eq{eq:axionpionLO}. 
Note that since we are interested in $2 \to 2$ scattering processes, we can neglect the  $\mathcal{O}(p^4)$ Wess-Zumino-Witten term \cite{Wess:1971yu,Witten:1983tw} since it contains operators with an odd number of bosons. 

To compute the axial current $J^{A}_\mu$ at NLO, we promote the ordinary derivative to a covariant one, defined as $\Dmu U = \dmu U -i r_\mu U + i U l_\mu$, with $r_\mu=r_\mu^A \sigma^A/2$ and $l_\mu=l_\mu^A \sigma^A/2$ external fields which can be used to include electromagnetic or weak effects. 
The left and right SU(2) currents are obtained by differentiating the NLO Lagrangian with respect to  $l_\mu^A$ and $r_\mu^A$, respectively. Taking the $R-L$ combination and switching off the external fields, the NLO axial current reads
\begin{align}
\label{eq:NLOCurrent}
&J^{A}_\mu|^{\rm NLO}=
    \frac{i}{2} \ell_1 {\rm Tr}\left[ \sigma^A \left\{\dmu\Ud, U\right\}\right] {\rm Tr}\left[ \dnu U \dnuu\Ud \right] \nonumber\\
    &+\frac{i}{4} \ell_2 {\rm Tr}\left[ \sigma^A \left\{\dnuu\Ud, U \right\} \right]{\rm Tr}\left[ \dmu U \dnu\Ud + \dnu U \dmu \Ud\right]\nonumber\\
    &-\frac{i}{8} \ell_4 {\rm Tr} \big[ \sigma^A \left\{\dmu U, \Chid_a\right\}- \sigma^A \left\{U, \dmu \Chid_a \right\} \nonumber\\
    &\ \ \ \ \ \ \ +\sigma^A \left\{\dmu \chi_a,\Ud\right\}-\sigma^A\left\{\chi_a,\dmu \Ud\right\} \big] \, ,
    \end{align}
 where curly brackets indicate anti-commutators.
 
New $a\text{-}\piz$ mixings arise at NLO, both at tree level from the NLO Lagrangian 
and at one loop from $\mathscr{L}^{\rm LO}_{a\text{-}\pi}$. 
These mixings are explicitly taken into account in the 
Lehmann-Symanzik-Zimmermann (LSZ) reduction formula \cite{Lehmann:1954rq} 
(focussing e.g.~on the $a\piz \to \pip \pim$ channel)
\begin{align}
\mathcal{M}_{a\piz \to \pip \pim}&=\frac{1}{\sqrt{Z_a Z_\pi^3}} \Pi_{i=1}^{4} \lim_{p_i^2 \to m_i^2} \left(p_i^2-m_i^2\right) \nonumber \\
&\times G_{a\piz\pip\pim} (p_1,p_2,p_3,p_4) \, , 
\end{align}
where the index $i$ runs over the external particles,  
$Z_a$ ($Z_\pi$) is the wave-function renormalization 
of the axion (pion) field and  
the full 4-point Green's function 
is given by
\begin{align}
 \label{GreenF}
&G_{a\piz\pip\pim}  =\sum_{i,j=a,\piz}\G_{i j \pip \pim } \\
&\times G_{\pip\pip}(m^2_{\pi}) G_{\pim\pim}(m^2_{\pi}) G_{a i}(m^2_{a} = 0) G_{\piz j}(m^2_{\pi}) \, .
\nonumber
\end{align}
The first term is the amputated 4-point function, multiplied by the 2-point 
functions of the external legs with the axion mass to zero. 
Working with LO diagonal propagators, the 2-point amplitude 
for the $a\text{-}\piz$ system reads 
$\P_{ij} = \diag(p^2, p^2-m^2_\pi) - \Sigma_{ij}$,  
where $\Sigma_{ij}$ encodes NLO corrections including mixings. 
The 2-point Green's function $G_{ij} =(-i \P)^{-1}_{ij}$
is hence 
\beq
\label{2points}
G_{ij} 
=i \begin{pmatrix} 
\frac{1}{p^2} &\frac{\Sigma_{a\pi}}{p^2\(p^2-m_\pi^2-\Sigma_{\pi\pi}\)}\\
\frac{\Sigma_{a\pi}}{p^2\(p^2-m_\pi^2-\Sigma_{\pi\pi}\)} & \frac{1}{p^2-m_\pi^2-\Sigma_{\pi\pi}}
\end{pmatrix} \, . 
\eeq
Plugging \eq{GreenF} and (\ref{2points}) into the LSZ formula for the scattering amplitude and neglecting $\mathcal{O}(1/f_a)^2$ terms, 
one finds
(with $Z_a = 1$, $Z_\pi = 1 + \Sigma'_{\pi\pi} (m^2_\pi)$ 
and primes indicating derivatives with respect to $p^2$)
\begin{align}
&\mathcal{M}_{a\piz \to \pip \pim}
=\(1+\frac{3}{2} \Sigma'_{\pi\pi}
(m^2_\pi)
\) 
\G_{a\piz\pip\pim}^{\rm LO}  \nonumber \\
& 
-\frac{\Sigma_{a\pi}
(m_a^2= 0)
}{m^2_\pi}\G_{\piz\piz\pip\pim}^{\rm LO} 
+ \G_{a\piz\pip\pim}^{\rm NLO} \, , 
\end{align}
where the 
$\G$'s
are evaluated at the physical masses of the external particles. 
The one-loop amplitudes have been computed in dimensional regularization.  
To carry out the renormalization procedure in the (modified) $\overline{\text{MS}}$ scheme,
we define the scale independent parameters $\overline{\ell_i}$ as \cite{Gasser:1983yg}
\beq
\label{lbar}
        \ell_i=\frac{\gamma_i}{32\pi^2}\left[\overline{\ell_i}+R+\ln\left(\frac{m_\pi^2}{\mu^2}\right) \right] \, , 
\eeq
with $R= \frac{2}{d-4}- \log(4\pi)+\gamma_E-1$, in order to cancel the divergent terms (in the limit $d=4$) with a suitable choice of the $\gamma_i$. 
Eventually, only the terms proportional to
$\ell_{1,2,7}$ contribute to the NLO amplitude, 
which is renormalized for $\gamma_1=1/3$, $\gamma_2=2/3$ and $\gamma_7=0$. 
The latter coincide with the values obtained in Ref.~\cite{Gasser:1983yg} for 
the standard chiral theory without the axion. 
	
The renormalized NLO amplitude for the $a \piz\rightarrow\pip \pim$ process
(and its crossed channels)  
is given in \eq{eq:Mapi0pippimNLO} of
the Supplementary Material.
%, which includes Refs.~\cite{FeynRules1,feynRules2,feynArts,feynCalc1,feynCalc2,feynCalc3,Patel}. 
We have also checked that the same analytical result is obtained 
via a direct NLO diagonalization of the $a$ and $\pi^0$ propagators, 
without employing the LSZ formalism with off-diagonal propagators. 
For consistency, we will only consider the interference between 
the LO and NLO terms in the squared matrix elements,
$\sum |\mathcal{M}|^2\simeq 
\sum |\mathcal{M}|_{\rm LO}^2 +  \sum 2\Re[\mathcal{M}_{\rm LO} \mathcal{M}^*_{\rm NLO}]$, 
since the NLO squared correction 
is of the same order of the NNLO-LO interference, 
which we neglect. 

\Sec{Breakdown of the chiral expansion at finite temperature.}
The crucial quantity that is needed to 
extract the HDM bound 
is the axion decoupling temperature, $T_D$, 
obtained via the
freeze-out condition 
(following the same criterium as in \cite{Hannestad:2005df})
\begin{equation}
\label{eq:dectemp}
    \Gamma_a(T_D) = H(T_D) \, .
\end{equation}
Here, $H(T) = \sqrt{4\pi^3 g_\star(T) / 45} \, T^2 / m_{\rm pl}$ 
is the Hubble rate (assuming a radiation dominated Universe)
in terms of the Planck mass $m_{\rm pl} = 1.22 \times 10^{19}$ GeV 
and the effective number of 
relativistic degrees of freedom, 
$g_\star(T)$, while 
$\Gamma_a$ is the axion thermalization rate 
entering the Boltzmann equation
\begin{align}\label{Gamma1}
\Gamma_a 
&= \frac{1}{n_a^{\rm eq}} \int\frac{d^3 \mathbf{p}_1}{(2\pi)^3 2 E_1}\frac{d^3  \mathbf{p}_2}{(2\pi)^3 2 E_2}\frac{d^3  \mathbf{p}_3}{(2\pi)^3 2 E_3}\frac{d^3  \mathbf{p}_4}{(2\pi)^3 2 E_4}  
\nonumber \\
&\times  \sum |\mathcal{M}|^2 (2\pi)^4 \delta^4
\left( 
p_1+p_2-p_3-p_4
\right) \nonumber \\
&\times  f_1 f_2 
(1+ f_3)(1+ f_4) \, , 
\end{align}
where $n_a^{\rm eq} = (\zeta_3 / \pi^2) T^3$ 
and $f_{i} = 1/(e^{E_{i}/T} - 1)$.  

In the following, we will set 
the model-dependent axion couplings
$c^0_{u,\,d}=0$ (cf.~\eq{eq:defCapi}),   
to comply with  
the standard setup considered in 
the literature \cite{Chang:1993gm,Hannestad:2005df,Melchiorri:2007cd,Hannestad:2008js,Hannestad:2010yi,Archidiacono:2013cha,Giusarma:2014zza,DiValentino:2015zta,DiValentino:2015wba,Archidiacono:2015mda,Giare:2020vzo} (see \cite{Ferreira:2020bpb} for an exception). 
Moreover, we will neglect thermal corrections to the scattering matrix element,  
since those are small for $T \lesssim m_\pi$ \cite{Gasser:1986vb,Gasser:1987ah,Gerber:1988tt}. 
%\begin{figure}[t]
%\centering
%\includegraphics[width=8cm]{h_functions.pdf}
%\vspace*{-4ex}%
%\caption{Numerical profile of the $h_{\rm LO}$ and $h_{\rm NLO}$ functions 
%entering the axion-pion thermalization rate in \eq{gammaInterf}.}
%\label{fig:hfuncts}       
%\end{figure}
By integrating numerically the phase space in \eq{Gamma1}
we find (see \eq{GammaInt} of the Supplementary Material 
for a useful intermediate analytical step, or \cite{Hannestad:1995rs} for a slightly different approach)
\begin{align}\label{gammaInterf}
  \Gamma_a(T) &= \left(  \frac{C_{a\pi}}{f_a f_\pi}\right)^2 
   0.212 \ T^5 \Big[ h_{\rm LO}(m_\pi/T)   \nonumber \\ 
  & - 2.92 \frac{T^2}{f_\pi^2}\ h_{\rm NLO}(m_\pi/T)\Big] \, , 
\end{align}
where for the numerical evaluation we used the 
central values of the
LECs 
$\overline{\ell_1} = -0.36(59)$ \cite{Colangelo:2001df}, 
$\overline{\ell_2} = 4.31(11)$ \cite{Colangelo:2001df},
$\overline{\ell_3} = 3.53(26)$ \cite{Aoki:2019cca},
$\overline{\ell_4} = 4.73(10)$ \cite{Aoki:2019cca}
and 
$\ell_7 = 7(4) \times 10^{-3}$ \cite{diCortona:2015ldu}, 
$m_u / m_d = 0.50(2)$ \cite{Aoki:2019cca}, 
$f_\pi = 92.1(8)$ MeV \cite{Zyla:2020zbs} and $m_\pi = 137$ MeV 
(corresponding to the average neutral/charged pion mass). 
The $h$-functions are normalized to $h_{\rm LO}(0)=h_{\rm NLO}(0)=1$ 
and 
%their numerical profile is reported 
they are plotted in \fig{fig:hfuncts}
of the Supplementary Material.  
We have checked that $h_{\rm LO}$ reproduces 
the result 
of Ref.~\cite{Hannestad:2005df} 
within percent accuracy. 
It should be noted that 
\eq{gammaInterf} 
is meaningful only for $m_\pi / T \gtrsim 1$, since at 
higher temperatures above $T_c$ 
pions are deconfined and the axion thermalization rate 
should be computed from the interactions with a quark-gluon plasma. 
Nevertheless, we are interested in extrapolating the behaviour 
of \eq{gammaInterf} from the low-temperature regime, 
where the chiral approach is reliable.  

\begin{figure}[t]
\centering
\includegraphics[width=8cm]{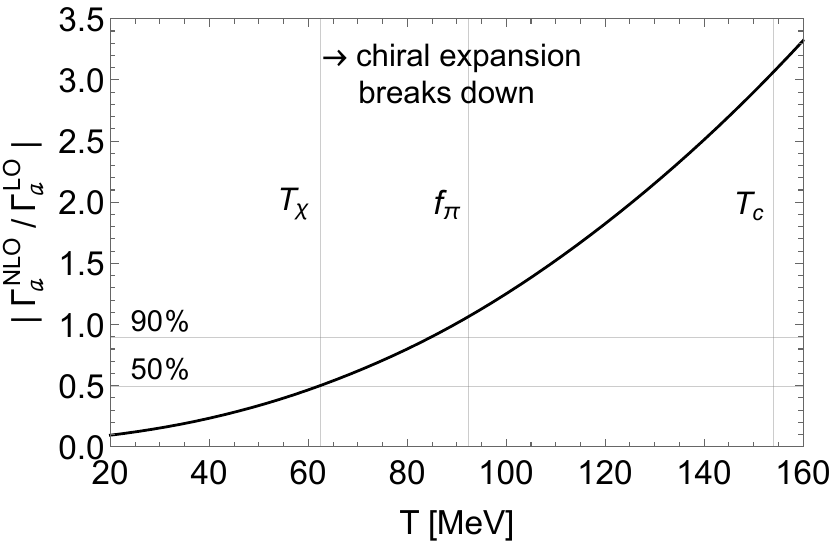} 
\vspace*{-4ex}%
\caption{Ratio between the NLO and the LO axion-pion thermalization rate. 
$T_\chi \simeq 62$ MeV corresponds to a NLO correction of $50\%$. }
\label{fig:NLOvsLO}       
\end{figure}

In \fig{fig:NLOvsLO} we compare the LO and NLO rates contributing to 
$\Gamma_a = \Gamma_a^{\rm LO} + \Gamma_a^{\rm NLO}$. 
In particular, the $|\Gamma_a^{\rm NLO} / \Gamma_a^{\rm LO}|$ ratio
does not depend on $f_a$.  
Requiring as a loose criterium 
that the NLO correction is less than $50\%$ 
of the LO one, yields $T_\chi \simeq 62$ MeV as the maximal temperature at which  
the chiral description of the thermalization rate
can be reliably extended. 

\begin{figure}[t]
\centering
\includegraphics[width=8cm]{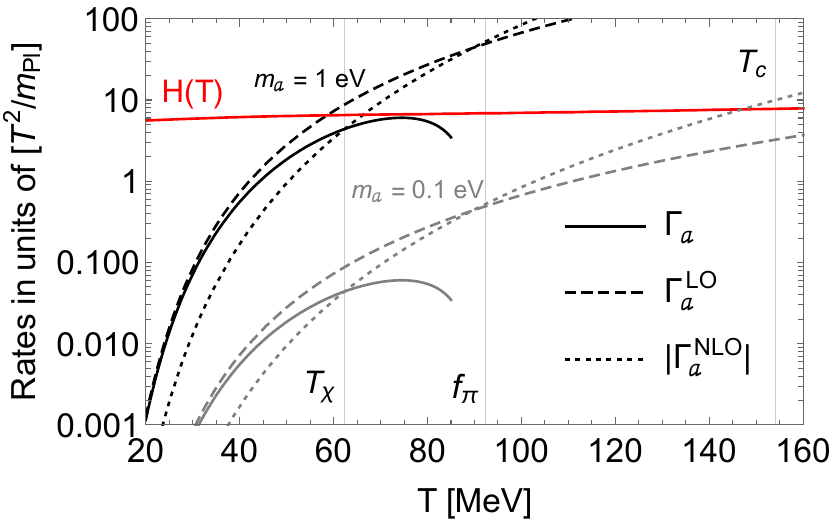}
\vspace*{-4ex}%
\caption{Axion-pion thermalization rate vs.~Hubble rate for two reference values 
of the axion mass, $m_a = 1$ eV and 0.1 eV. 
The full $\Gamma_a$ has been extrapolated for illustrative purposes until $T \simeq 85$ MeV, for which 
$|\Gamma^{\rm NLO}_a / \Gamma^{\rm LO}_a| = 90\%$. 
%Note, however, that the chiral expansion is supposed to fail earlier, around $T_\chi$.
}
\label{fig:Gammas}       
\end{figure}

\begin{figure}[t!]
\centering
\includegraphics[width=7.9cm]{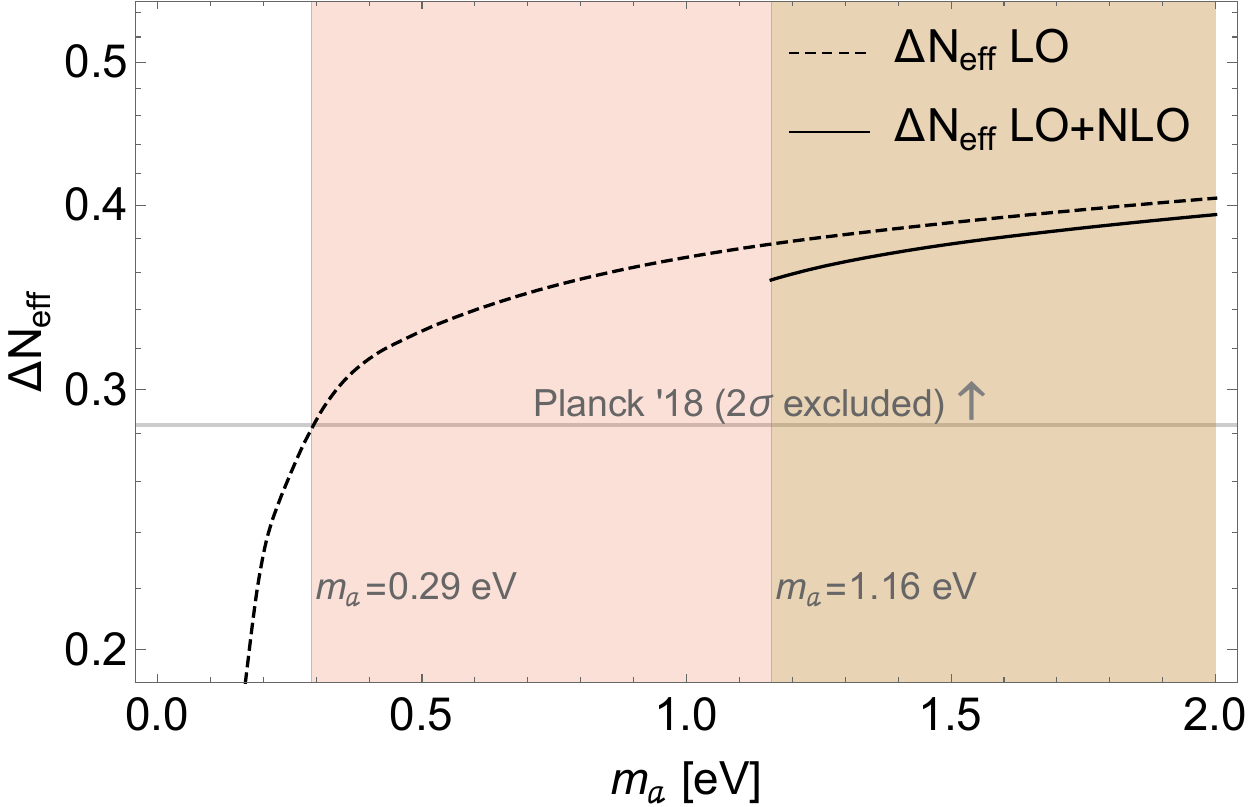}
\vspace*{-4ex}%
\caption{$\Delta N_{\rm eff}$ as a function of $m_a$. 
The LO+NLO prediction is truncated where the chiral expansion fails.}
\label{fig:DeltaNeff}       
\end{figure}

\fig{fig:Gammas} shows instead the extraction of the decoupling 
temperature (defined via \eq{eq:dectemp}) 
for two reference values of the axion mass 
(setting the strength of the axion coupling via $f_a$), 
namely $m_a = 1$ eV and 0.1 eV. 
Assuming a standard analysis employing the 
LO axion thermalization rate 
\cite{Hannestad:2005df}, 
the former benchmark (1 eV) corresponds to the most conservative HDM bound \cite{Giare:2020vzo}, 
while the latter (0.1 eV) saturates the most stringent one \cite{Giare:2020vzo}
and also represents the typical reach of future CMB-S4 experiments \cite{Abazajian:2016yjj}. 
However, from \fig{fig:Gammas} we see that 
$T_D^{\rm LO} \simeq 59$ MeV for $m_a = 1$ eV 
and 
$T_D^{\rm LO} \sim 200$ MeV for $m_a = 0.1$ eV. 
While in the former case the decoupling temperature 
is at the boundary of validity of the 
chiral expansion, set by $T_\chi \simeq 62$ MeV, 
in the latter 
is well above it. 
Hence, the region where the chiral expansion fails, 
$T_D \gtrsim T_\chi$, corresponds to $m_a \lesssim 1.2$ eV.   
Since $m_a \simeq 1.2$ eV yields 
a too large contribution to $\Delta N_{\rm eff}$ incompatible 
with Planck data (cf.~\fig{fig:DeltaNeff}),  
%\cite{Akrami:2018vks,Aghanim:2018eyx}, 
this value 
can be regarded as   
the axion HDM bound that can be reliably extracted within chiral perturbation theory.
However, 
in the relevant mass range $m_a \in [0.1,1]$ eV 
the decoupling temperature 
and consequently the axion HDM bound 
cannot be reliably extracted within the chiral approach. 
%This is explicitly shown in \fig{fig:DeltaNeff}. 

Note, finally, that in the presence of 
model-dependent axion couplings 
$c^0_{u,d} \gg 1$ (as in some 
axion models \cite{Darme:2020gyx}), 
the same decoupling temperature as in the $c^0_{u,d} = 0$ case is obtained for larger 
$f_a$, 
thus shifting down the mass window relevant for the axion HDM bound.  

\Sec{Towards a reliable axion HDM bound.}  
The failure of the chiral approach in the calculation of the 
axion-pion thermalization rate 
can be traced back to the fact that 
in a thermal bath with temperatures of the order of $T \simeq 100$ MeV 
the mean energy of pions is $\vev{E_\pi} \simeq 350$ MeV, 
so that $\pi$-$\pi$ scatterings happen at center of mass energies 
above the validity of the 2-flavour chiral EFT. 
The latter can be related to the scale of tree-level unitarity violation 
of $\pi$-$\pi$ scattering resulting in 
$\sqrt{s} \lesssim \sqrt{8 \pi} f_\pi \simeq 460$ MeV \cite{Weinberg:1966kf,Aydemir:2012nz}. 
A possible strategy to extend the theoretical predictions at higher energies is to compute the
relevant  $a\pi \to \pi\pi$ amplitudes using  lattice QCD simulations.  To this end one may employ 
the standard techniques used to compute weak non-leptonic  matrix elements~\cite{Blum:2015ywa,Abbott:2020hxn} and $\pi$-$\pi$ scattering amplitudes as a function of the energy  at finite volume~\cite{Luscher:1990ux, Rummukainen:1995vs,Kim:2005gf,Hansen:2012tf}. Although this approach has limitations with respect to the maximum attainable  center of mass energy,   we believe that it  can be used to compute the amplitudes up to values of $\sqrt{s}\sim 600-900$~MeV or higher~\cite{Briceno:2016mjc}.

We conclude by stressing the importance of obtaining a reliable determination of the 
axion-pion thermalization rate, not only in view of the extraction of a 
notable bound in axion physics, but also in order to 
set definite targets for future CMB probes of the axion-pion coupling, 
which could represent a `discovery channel' for the axion.

\begin{acknowledgments}
\Sec{Acknowledgments.} 
We thank Enrico Nardi and Maurizio Giannotti for helpful discussions. 
The work of LDL is supported by the Marie Sk\l{}odowska-Curie 
Individual Fellowship grant AXIONRUSH (GA 840791) 
and the Deutsche Forschungsgemeinschaft under Germany's Excellence Strategy 
- EXC 2121 Quantum Universe - 390833306. The work of GP has received funding from the European Union's Horizon 2020 research and innovation programme under the 
Marie Sk\l{}odowska-Curie grant agreement N$^{\circ}$ 860881. 
\end{acknowledgments}

\bibliographystyle{apsrev4-1.bst}
\bibliography{bibliography}

\begin{thebibliography}{81}%
\makeatletter
\providecommand \@ifxundefined [1]{%
 \@ifx{#1\undefined}
}%
\providecommand \@ifnum [1]{%
 \ifnum #1\expandafter \@firstoftwo
 \else \expandafter \@secondoftwo
 \fi
}%
\providecommand \@ifx [1]{%
 \ifx #1\expandafter \@firstoftwo
 \else \expandafter \@secondoftwo
 \fi
}%
\providecommand \natexlab [1]{#1}%
\providecommand \enquote  [1]{``#1''}%
\providecommand \bibnamefont  [1]{#1}%
\providecommand \bibfnamefont [1]{#1}%
\providecommand \citenamefont [1]{#1}%
\providecommand \href@noop [0]{\@secondoftwo}%
\providecommand \href [0]{\begingroup \@sanitize@url \@href}%
\providecommand \@href[1]{\@@startlink{#1}\@@href}%
\providecommand \@@href[1]{\endgroup#1\@@endlink}%
\providecommand \@sanitize@url [0]{\catcode `\\12\catcode `\$12\catcode
  `\&12\catcode `\#12\catcode `\^12\catcode `\_12\catcode `\%12\relax}%
\providecommand \@@startlink[1]{}%
\providecommand \@@endlink[0]{}%
\providecommand \url  [0]{\begingroup\@sanitize@url \@url }%
\providecommand \@url [1]{\endgroup\@href {#1}{\urlprefix }}%
\providecommand \urlprefix  [0]{URL }%
\providecommand \Eprint [0]{\href }%
\providecommand \doibase [0]{http://dx.doi.org/}%
\providecommand \selectlanguage [0]{\@gobble}%
\providecommand \bibinfo  [0]{\@secondoftwo}%
\providecommand \bibfield  [0]{\@secondoftwo}%
\providecommand \translation [1]{[#1]}%
\providecommand \BibitemOpen [0]{}%
\providecommand \bibitemStop [0]{}%
\providecommand \bibitemNoStop [0]{.\EOS\space}%
\providecommand \EOS [0]{\spacefactor3000\relax}%
\providecommand \BibitemShut  [1]{\csname bibitem#1\endcsname}%
\let\auto@bib@innerbib\@empty
%</preamble>
\bibitem [{\citenamefont {Peccei}\ and\ \citenamefont
  {Quinn}(1977{\natexlab{a}})}]{Peccei:1977hh}%
  \BibitemOpen
  \bibfield  {author} {\bibinfo {author} {\bibfnamefont {R.~D.}\ \bibnamefont
  {Peccei}}\ and\ \bibinfo {author} {\bibfnamefont {H.~R.}\ \bibnamefont
  {Quinn}},\ }\href {\doibase 10.1103/PhysRevLett.38.1440} {\bibfield
  {journal} {\bibinfo  {journal} {Phys. Rev. Lett.}\ }\textbf {\bibinfo
  {volume} {38}},\ \bibinfo {pages} {1440} (\bibinfo {year}
  {1977}{\natexlab{a}})}\BibitemShut {NoStop}%
%%CITATION = PRLTA,38,1440;%%
\bibitem [{\citenamefont {Peccei}\ and\ \citenamefont
  {Quinn}(1977{\natexlab{b}})}]{Peccei:1977ur}%
  \BibitemOpen
  \bibfield  {author} {\bibinfo {author} {\bibfnamefont {R.~D.}\ \bibnamefont
  {Peccei}}\ and\ \bibinfo {author} {\bibfnamefont {H.~R.}\ \bibnamefont
  {Quinn}},\ }\href {\doibase 10.1103/PhysRevD.16.1791} {\bibfield  {journal}
  {\bibinfo  {journal} {Phys. Rev.}\ }\textbf {\bibinfo {volume} {D16}},\
  \bibinfo {pages} {1791} (\bibinfo {year} {1977}{\natexlab{b}})}\BibitemShut
  {NoStop}%
%%CITATION = PHRVA,D16,1791;%%
\bibitem [{\citenamefont {Wilczek}(1978)}]{Wilczek:1977pj}%
  \BibitemOpen
  \bibfield  {author} {\bibinfo {author} {\bibfnamefont {F.}~\bibnamefont
  {Wilczek}},\ }\href {\doibase 10.1103/PhysRevLett.40.279} {\bibfield
  {journal} {\bibinfo  {journal} {Phys. Rev. Lett.}\ }\textbf {\bibinfo
  {volume} {40}},\ \bibinfo {pages} {279} (\bibinfo {year} {1978})}\BibitemShut
  {NoStop}%
%%CITATION = PRLTA,40,279;%%
\bibitem [{\citenamefont {Weinberg}(1978)}]{Weinberg:1977ma}%
  \BibitemOpen
  \bibfield  {author} {\bibinfo {author} {\bibfnamefont {S.}~\bibnamefont
  {Weinberg}},\ }\href {\doibase 10.1103/PhysRevLett.40.223} {\bibfield
  {journal} {\bibinfo  {journal} {Phys. Rev. Lett.}\ }\textbf {\bibinfo
  {volume} {40}},\ \bibinfo {pages} {223} (\bibinfo {year} {1978})}\BibitemShut
  {NoStop}%
%%CITATION = PRLTA,40,223;%%
\bibitem [{\citenamefont {Preskill}\ \emph {et~al.}(1983)\citenamefont
  {Preskill}, \citenamefont {Wise},\ and\ \citenamefont
  {Wilczek}}]{Preskill:1982cy}%
  \BibitemOpen
  \bibfield  {author} {\bibinfo {author} {\bibfnamefont {J.}~\bibnamefont
  {Preskill}}, \bibinfo {author} {\bibfnamefont {M.~B.}\ \bibnamefont {Wise}},
  \ and\ \bibinfo {author} {\bibfnamefont {F.}~\bibnamefont {Wilczek}},\ }\href
  {\doibase 10.1016/0370-2693(83)90637-8} {\bibfield  {journal} {\bibinfo
  {journal} {Phys. Lett.}\ }\textbf {\bibinfo {volume} {120B}},\ \bibinfo
  {pages} {127} (\bibinfo {year} {1983})}\BibitemShut {NoStop}%
%%CITATION = PHLTA,120B,127;%%
\bibitem [{\citenamefont {Abbott}\ and\ \citenamefont
  {Sikivie}(1983)}]{Abbott:1982af}%
  \BibitemOpen
  \bibfield  {author} {\bibinfo {author} {\bibfnamefont {L.~F.}\ \bibnamefont
  {Abbott}}\ and\ \bibinfo {author} {\bibfnamefont {P.}~\bibnamefont
  {Sikivie}},\ }\href {\doibase 10.1016/0370-2693(83)90638-X} {\bibfield
  {journal} {\bibinfo  {journal} {Phys. Lett.}\ }\textbf {\bibinfo {volume}
  {120B}},\ \bibinfo {pages} {133} (\bibinfo {year} {1983})}\BibitemShut
  {NoStop}%
%%CITATION = PHLTA,120B,133;%%
\bibitem [{\citenamefont {Dine}\ and\ \citenamefont
  {Fischler}(1983)}]{Dine:1982ah}%
  \BibitemOpen
  \bibfield  {author} {\bibinfo {author} {\bibfnamefont {M.}~\bibnamefont
  {Dine}}\ and\ \bibinfo {author} {\bibfnamefont {W.}~\bibnamefont
  {Fischler}},\ }\href {\doibase 10.1016/0370-2693(83)90639-1} {\bibfield
  {journal} {\bibinfo  {journal} {Phys. Lett.}\ }\textbf {\bibinfo {volume}
  {120B}},\ \bibinfo {pages} {137} (\bibinfo {year} {1983})}\BibitemShut
  {NoStop}%
%%CITATION = PHLTA,120B,137;%%
\bibitem [{\citenamefont {Davis}(1986)}]{Davis:1986xc}%
  \BibitemOpen
  \bibfield  {author} {\bibinfo {author} {\bibfnamefont {R.~L.}\ \bibnamefont
  {Davis}},\ }\href {\doibase 10.1016/0370-2693(86)90300-X} {\bibfield
  {journal} {\bibinfo  {journal} {Phys. Lett. B}\ }\textbf {\bibinfo {volume}
  {180}},\ \bibinfo {pages} {225} (\bibinfo {year} {1986})}\BibitemShut
  {NoStop}%
\bibitem [{\citenamefont {Turner}(1987)}]{Turner:1986tb}%
  \BibitemOpen
  \bibfield  {author} {\bibinfo {author} {\bibfnamefont {M.~S.}\ \bibnamefont
  {Turner}},\ }\href {\doibase 10.1103/PhysRevLett.59.2489} {\bibfield
  {journal} {\bibinfo  {journal} {Phys. Rev. Lett.}\ }\textbf {\bibinfo
  {volume} {59}},\ \bibinfo {pages} {2489} (\bibinfo {year} {1987})},\ \bibinfo
  {note} {[Erratum: Phys.Rev.Lett. 60, 1101 (1988)]}\BibitemShut {NoStop}%
\bibitem [{\citenamefont {Kolb}\ and\ \citenamefont
  {Turner}(1990)}]{Kolb:1990vq}%
  \BibitemOpen
  \bibfield  {author} {\bibinfo {author} {\bibfnamefont {E.~W.}\ \bibnamefont
  {Kolb}}\ and\ \bibinfo {author} {\bibfnamefont {M.~S.}\ \bibnamefont
  {Turner}},\ }\href@noop {} {\emph {\bibinfo {title} {{The Early
  Universe}}}},\ Vol.~\bibinfo {volume} {69}\ (\bibinfo {year}
  {1990})\BibitemShut {NoStop}%
\bibitem [{\citenamefont {Aghanim}\ \emph
  {et~al.}(2020{\natexlab{a}})\citenamefont {Aghanim} \emph
  {et~al.}}]{Akrami:2018vks}%
  \BibitemOpen
  \bibfield  {author} {\bibinfo {author} {\bibfnamefont {N.}~\bibnamefont
  {Aghanim}} \emph {et~al.} (\bibinfo {collaboration} {Planck}),\ }\href
  {\doibase 10.1051/0004-6361/201833880} {\bibfield  {journal} {\bibinfo
  {journal} {Astron. Astrophys.}\ }\textbf {\bibinfo {volume} {641}},\ \bibinfo
  {pages} {A1} (\bibinfo {year} {2020}{\natexlab{a}})},\ \Eprint
  {http://arxiv.org/abs/1807.06205} {arXiv:1807.06205 [astro-ph.CO]}
  \BibitemShut {NoStop}%
\bibitem [{\citenamefont {Aghanim}\ \emph
  {et~al.}(2020{\natexlab{b}})\citenamefont {Aghanim} \emph
  {et~al.}}]{Aghanim:2018eyx}%
  \BibitemOpen
  \bibfield  {author} {\bibinfo {author} {\bibfnamefont {N.}~\bibnamefont
  {Aghanim}} \emph {et~al.} (\bibinfo {collaboration} {Planck}),\ }\href
  {\doibase 10.1051/0004-6361/201833910} {\bibfield  {journal} {\bibinfo
  {journal} {Astron. Astrophys.}\ }\textbf {\bibinfo {volume} {641}},\ \bibinfo
  {pages} {A6} (\bibinfo {year} {2020}{\natexlab{b}})},\ \Eprint
  {http://arxiv.org/abs/1807.06209} {arXiv:1807.06209 [astro-ph.CO]}
  \BibitemShut {NoStop}%
\bibitem [{\citenamefont {Abazajian}\ \emph {et~al.}(2016)\citenamefont
  {Abazajian} \emph {et~al.}}]{Abazajian:2016yjj}%
  \BibitemOpen
  \bibfield  {author} {\bibinfo {author} {\bibfnamefont {K.~N.}\ \bibnamefont
  {Abazajian}} \emph {et~al.} (\bibinfo {collaboration} {CMB-S4}),\ }\href@noop
  {} {\  (\bibinfo {year} {2016})},\ \Eprint {http://arxiv.org/abs/1610.02743}
  {arXiv:1610.02743 [astro-ph.CO]} \BibitemShut {NoStop}%
\bibitem [{\citenamefont {Salvio}\ \emph {et~al.}(2014)\citenamefont {Salvio},
  \citenamefont {Strumia},\ and\ \citenamefont {Xue}}]{Salvio:2013iaa}%
  \BibitemOpen
  \bibfield  {author} {\bibinfo {author} {\bibfnamefont {A.}~\bibnamefont
  {Salvio}}, \bibinfo {author} {\bibfnamefont {A.}~\bibnamefont {Strumia}}, \
  and\ \bibinfo {author} {\bibfnamefont {W.}~\bibnamefont {Xue}},\ }\href
  {\doibase 10.1088/1475-7516/2014/01/011} {\bibfield  {journal} {\bibinfo
  {journal} {JCAP}\ }\textbf {\bibinfo {volume} {01}},\ \bibinfo {pages} {011}
  (\bibinfo {year} {2014})},\ \Eprint {http://arxiv.org/abs/1310.6982}
  {arXiv:1310.6982 [hep-ph]} \BibitemShut {NoStop}%
\bibitem [{\citenamefont {Baumann}\ \emph {et~al.}(2016)\citenamefont
  {Baumann}, \citenamefont {Green},\ and\ \citenamefont
  {Wallisch}}]{Baumann:2016wac}%
  \BibitemOpen
  \bibfield  {author} {\bibinfo {author} {\bibfnamefont {D.}~\bibnamefont
  {Baumann}}, \bibinfo {author} {\bibfnamefont {D.}~\bibnamefont {Green}}, \
  and\ \bibinfo {author} {\bibfnamefont {B.}~\bibnamefont {Wallisch}},\ }\href
  {\doibase 10.1103/PhysRevLett.117.171301} {\bibfield  {journal} {\bibinfo
  {journal} {Phys. Rev. Lett.}\ }\textbf {\bibinfo {volume} {117}},\ \bibinfo
  {pages} {171301} (\bibinfo {year} {2016})},\ \Eprint
  {http://arxiv.org/abs/1604.08614} {arXiv:1604.08614 [astro-ph.CO]}
  \BibitemShut {NoStop}%
\bibitem [{\citenamefont {Ferreira}\ and\ \citenamefont
  {Notari}(2018)}]{Ferreira:2018vjj}%
  \BibitemOpen
  \bibfield  {author} {\bibinfo {author} {\bibfnamefont {R.~Z.}\ \bibnamefont
  {Ferreira}}\ and\ \bibinfo {author} {\bibfnamefont {A.}~\bibnamefont
  {Notari}},\ }\href {\doibase 10.1103/PhysRevLett.120.191301} {\bibfield
  {journal} {\bibinfo  {journal} {Phys. Rev. Lett.}\ }\textbf {\bibinfo
  {volume} {120}},\ \bibinfo {pages} {191301} (\bibinfo {year} {2018})},\
  \Eprint {http://arxiv.org/abs/1801.06090} {arXiv:1801.06090 [hep-ph]}
  \BibitemShut {NoStop}%
\bibitem [{\citenamefont {Arias-Aragon}\ \emph {et~al.}(2020)\citenamefont
  {Arias-Aragon}, \citenamefont {D'Eramo}, \citenamefont {Ferreira},
  \citenamefont {Merlo},\ and\ \citenamefont {Notari}}]{Arias-Aragon:2020shv}%
  \BibitemOpen
  \bibfield  {author} {\bibinfo {author} {\bibfnamefont {F.}~\bibnamefont
  {Arias-Aragon}}, \bibinfo {author} {\bibfnamefont {F.}~\bibnamefont
  {D'Eramo}}, \bibinfo {author} {\bibfnamefont {R.~Z.}\ \bibnamefont
  {Ferreira}}, \bibinfo {author} {\bibfnamefont {L.}~\bibnamefont {Merlo}}, \
  and\ \bibinfo {author} {\bibfnamefont {A.}~\bibnamefont {Notari}},\
  }\href@noop {} {\  (\bibinfo {year} {2020})},\ \Eprint
  {http://arxiv.org/abs/2012.04736} {arXiv:2012.04736 [hep-ph]} \BibitemShut
  {NoStop}%
\bibitem [{\citenamefont {D'Eramo}\ \emph {et~al.}(2018)\citenamefont
  {D'Eramo}, \citenamefont {Ferreira}, \citenamefont {Notari},\ and\
  \citenamefont {Bernal}}]{DEramo:2018vss}%
  \BibitemOpen
  \bibfield  {author} {\bibinfo {author} {\bibfnamefont {F.}~\bibnamefont
  {D'Eramo}}, \bibinfo {author} {\bibfnamefont {R.~Z.}\ \bibnamefont
  {Ferreira}}, \bibinfo {author} {\bibfnamefont {A.}~\bibnamefont {Notari}}, \
  and\ \bibinfo {author} {\bibfnamefont {J.~L.}\ \bibnamefont {Bernal}},\
  }\href {\doibase 10.1088/1475-7516/2018/11/014} {\bibfield  {journal}
  {\bibinfo  {journal} {JCAP}\ }\textbf {\bibinfo {volume} {11}},\ \bibinfo
  {pages} {014} (\bibinfo {year} {2018})},\ \Eprint
  {http://arxiv.org/abs/1808.07430} {arXiv:1808.07430 [hep-ph]} \BibitemShut
  {NoStop}%
\bibitem [{\citenamefont {Masso}\ \emph {et~al.}(2002)\citenamefont {Masso},
  \citenamefont {Rota},\ and\ \citenamefont {Zsembinszki}}]{Masso:2002np}%
  \BibitemOpen
  \bibfield  {author} {\bibinfo {author} {\bibfnamefont {E.}~\bibnamefont
  {Masso}}, \bibinfo {author} {\bibfnamefont {F.}~\bibnamefont {Rota}}, \ and\
  \bibinfo {author} {\bibfnamefont {G.}~\bibnamefont {Zsembinszki}},\ }\href
  {\doibase 10.1103/PhysRevD.66.023004} {\bibfield  {journal} {\bibinfo
  {journal} {Phys. Rev. D}\ }\textbf {\bibinfo {volume} {66}},\ \bibinfo
  {pages} {023004} (\bibinfo {year} {2002})},\ \Eprint
  {http://arxiv.org/abs/hep-ph/0203221} {arXiv:hep-ph/0203221} \BibitemShut
  {NoStop}%
\bibitem [{\citenamefont {Graf}\ and\ \citenamefont
  {Steffen}(2011)}]{Graf:2010tv}%
  \BibitemOpen
  \bibfield  {author} {\bibinfo {author} {\bibfnamefont {P.}~\bibnamefont
  {Graf}}\ and\ \bibinfo {author} {\bibfnamefont {F.~D.}\ \bibnamefont
  {Steffen}},\ }\href {\doibase 10.1103/PhysRevD.83.075011} {\bibfield
  {journal} {\bibinfo  {journal} {Phys. Rev. D}\ }\textbf {\bibinfo {volume}
  {83}},\ \bibinfo {pages} {075011} (\bibinfo {year} {2011})},\ \Eprint
  {http://arxiv.org/abs/1008.4528} {arXiv:1008.4528 [hep-ph]} \BibitemShut
  {NoStop}%
\bibitem [{\citenamefont {Berezhiani}\ \emph {et~al.}(1992)\citenamefont
  {Berezhiani}, \citenamefont {Sakharov},\ and\ \citenamefont
  {Khlopov}}]{Berezhiani:1992rk}%
  \BibitemOpen
  \bibfield  {author} {\bibinfo {author} {\bibfnamefont {Z.}~\bibnamefont
  {Berezhiani}}, \bibinfo {author} {\bibfnamefont {A.}~\bibnamefont
  {Sakharov}}, \ and\ \bibinfo {author} {\bibfnamefont {M.}~\bibnamefont
  {Khlopov}},\ }\href@noop {} {\bibfield  {journal} {\bibinfo  {journal} {Sov.
  J. Nucl. Phys.}\ }\textbf {\bibinfo {volume} {55}},\ \bibinfo {pages} {1063}
  (\bibinfo {year} {1992})}\BibitemShut {NoStop}%
\bibitem [{\citenamefont {Chang}\ and\ \citenamefont
  {Choi}(1993)}]{Chang:1993gm}%
  \BibitemOpen
  \bibfield  {author} {\bibinfo {author} {\bibfnamefont {S.}~\bibnamefont
  {Chang}}\ and\ \bibinfo {author} {\bibfnamefont {K.}~\bibnamefont {Choi}},\
  }\href {\doibase 10.1016/0370-2693(93)90656-3} {\bibfield  {journal}
  {\bibinfo  {journal} {Phys. Lett. B}\ }\textbf {\bibinfo {volume} {316}},\
  \bibinfo {pages} {51} (\bibinfo {year} {1993})},\ \Eprint
  {http://arxiv.org/abs/hep-ph/9306216} {arXiv:hep-ph/9306216} \BibitemShut
  {NoStop}%
\bibitem [{\citenamefont {Hannestad}\ \emph {et~al.}(2005)\citenamefont
  {Hannestad}, \citenamefont {Mirizzi},\ and\ \citenamefont
  {Raffelt}}]{Hannestad:2005df}%
  \BibitemOpen
  \bibfield  {author} {\bibinfo {author} {\bibfnamefont {S.}~\bibnamefont
  {Hannestad}}, \bibinfo {author} {\bibfnamefont {A.}~\bibnamefont {Mirizzi}},
  \ and\ \bibinfo {author} {\bibfnamefont {G.}~\bibnamefont {Raffelt}},\ }\href
  {\doibase 10.1088/1475-7516/2005/07/002} {\bibfield  {journal} {\bibinfo
  {journal} {JCAP}\ }\textbf {\bibinfo {volume} {07}},\ \bibinfo {pages} {002}
  (\bibinfo {year} {2005})},\ \Eprint {http://arxiv.org/abs/hep-ph/0504059}
  {arXiv:hep-ph/0504059} \BibitemShut {NoStop}%
\bibitem [{\citenamefont {Zyla}\ \emph {et~al.}(2020)\citenamefont {Zyla} \emph
  {et~al.}}]{Zyla:2020zbs}%
  \BibitemOpen
  \bibfield  {author} {\bibinfo {author} {\bibfnamefont {P.}~\bibnamefont
  {Zyla}} \emph {et~al.} (\bibinfo {collaboration} {Particle Data Group}),\
  }\href {\doibase 10.1093/ptep/ptaa104} {\bibfield  {journal} {\bibinfo
  {journal} {PTEP}\ }\textbf {\bibinfo {volume} {2020}},\ \bibinfo {pages}
  {083C01} (\bibinfo {year} {2020})}\BibitemShut {NoStop}%
\bibitem [{\citenamefont {Melchiorri}\ \emph {et~al.}(2007)\citenamefont
  {Melchiorri}, \citenamefont {Mena},\ and\ \citenamefont
  {Slosar}}]{Melchiorri:2007cd}%
  \BibitemOpen
  \bibfield  {author} {\bibinfo {author} {\bibfnamefont {A.}~\bibnamefont
  {Melchiorri}}, \bibinfo {author} {\bibfnamefont {O.}~\bibnamefont {Mena}}, \
  and\ \bibinfo {author} {\bibfnamefont {A.}~\bibnamefont {Slosar}},\ }\href
  {\doibase 10.1103/PhysRevD.76.041303} {\bibfield  {journal} {\bibinfo
  {journal} {Phys. Rev. D}\ }\textbf {\bibinfo {volume} {76}},\ \bibinfo
  {pages} {041303} (\bibinfo {year} {2007})},\ \Eprint
  {http://arxiv.org/abs/0705.2695} {arXiv:0705.2695 [astro-ph]} \BibitemShut
  {NoStop}%
\bibitem [{\citenamefont {Hannestad}\ \emph {et~al.}(2008)\citenamefont
  {Hannestad}, \citenamefont {Mirizzi}, \citenamefont {Raffelt},\ and\
  \citenamefont {Wong}}]{Hannestad:2008js}%
  \BibitemOpen
  \bibfield  {author} {\bibinfo {author} {\bibfnamefont {S.}~\bibnamefont
  {Hannestad}}, \bibinfo {author} {\bibfnamefont {A.}~\bibnamefont {Mirizzi}},
  \bibinfo {author} {\bibfnamefont {G.~G.}\ \bibnamefont {Raffelt}}, \ and\
  \bibinfo {author} {\bibfnamefont {Y.~Y.}\ \bibnamefont {Wong}},\ }\href
  {\doibase 10.1088/1475-7516/2008/04/019} {\bibfield  {journal} {\bibinfo
  {journal} {JCAP}\ }\textbf {\bibinfo {volume} {04}},\ \bibinfo {pages} {019}
  (\bibinfo {year} {2008})},\ \Eprint {http://arxiv.org/abs/0803.1585}
  {arXiv:0803.1585 [astro-ph]} \BibitemShut {NoStop}%
\bibitem [{\citenamefont {Hannestad}\ \emph {et~al.}(2010)\citenamefont
  {Hannestad}, \citenamefont {Mirizzi}, \citenamefont {Raffelt},\ and\
  \citenamefont {Wong}}]{Hannestad:2010yi}%
  \BibitemOpen
  \bibfield  {author} {\bibinfo {author} {\bibfnamefont {S.}~\bibnamefont
  {Hannestad}}, \bibinfo {author} {\bibfnamefont {A.}~\bibnamefont {Mirizzi}},
  \bibinfo {author} {\bibfnamefont {G.~G.}\ \bibnamefont {Raffelt}}, \ and\
  \bibinfo {author} {\bibfnamefont {Y.~Y.}\ \bibnamefont {Wong}},\ }\href
  {\doibase 10.1088/1475-7516/2010/08/001} {\bibfield  {journal} {\bibinfo
  {journal} {JCAP}\ }\textbf {\bibinfo {volume} {08}},\ \bibinfo {pages} {001}
  (\bibinfo {year} {2010})},\ \Eprint {http://arxiv.org/abs/1004.0695}
  {arXiv:1004.0695 [astro-ph.CO]} \BibitemShut {NoStop}%
\bibitem [{\citenamefont {Archidiacono}\ \emph {et~al.}(2013)\citenamefont
  {Archidiacono}, \citenamefont {Hannestad}, \citenamefont {Mirizzi},
  \citenamefont {Raffelt},\ and\ \citenamefont {Wong}}]{Archidiacono:2013cha}%
  \BibitemOpen
  \bibfield  {author} {\bibinfo {author} {\bibfnamefont {M.}~\bibnamefont
  {Archidiacono}}, \bibinfo {author} {\bibfnamefont {S.}~\bibnamefont
  {Hannestad}}, \bibinfo {author} {\bibfnamefont {A.}~\bibnamefont {Mirizzi}},
  \bibinfo {author} {\bibfnamefont {G.}~\bibnamefont {Raffelt}}, \ and\
  \bibinfo {author} {\bibfnamefont {Y.~Y.}\ \bibnamefont {Wong}},\ }\href
  {\doibase 10.1088/1475-7516/2013/10/020} {\bibfield  {journal} {\bibinfo
  {journal} {JCAP}\ }\textbf {\bibinfo {volume} {10}},\ \bibinfo {pages} {020}
  (\bibinfo {year} {2013})},\ \Eprint {http://arxiv.org/abs/1307.0615}
  {arXiv:1307.0615 [astro-ph.CO]} \BibitemShut {NoStop}%
\bibitem [{\citenamefont {Giusarma}\ \emph {et~al.}(2014)\citenamefont
  {Giusarma}, \citenamefont {Di~Valentino}, \citenamefont {Lattanzi},
  \citenamefont {Melchiorri},\ and\ \citenamefont {Mena}}]{Giusarma:2014zza}%
  \BibitemOpen
  \bibfield  {author} {\bibinfo {author} {\bibfnamefont {E.}~\bibnamefont
  {Giusarma}}, \bibinfo {author} {\bibfnamefont {E.}~\bibnamefont
  {Di~Valentino}}, \bibinfo {author} {\bibfnamefont {M.}~\bibnamefont
  {Lattanzi}}, \bibinfo {author} {\bibfnamefont {A.}~\bibnamefont
  {Melchiorri}}, \ and\ \bibinfo {author} {\bibfnamefont {O.}~\bibnamefont
  {Mena}},\ }\href {\doibase 10.1103/PhysRevD.90.043507} {\bibfield  {journal}
  {\bibinfo  {journal} {Phys. Rev. D}\ }\textbf {\bibinfo {volume} {90}},\
  \bibinfo {pages} {043507} (\bibinfo {year} {2014})},\ \Eprint
  {http://arxiv.org/abs/1403.4852} {arXiv:1403.4852 [astro-ph.CO]} \BibitemShut
  {NoStop}%
\bibitem [{\citenamefont {Di~Valentino}\ \emph {et~al.}(2015)\citenamefont
  {Di~Valentino}, \citenamefont {Gariazzo}, \citenamefont {Giusarma},\ and\
  \citenamefont {Mena}}]{DiValentino:2015zta}%
  \BibitemOpen
  \bibfield  {author} {\bibinfo {author} {\bibfnamefont {E.}~\bibnamefont
  {Di~Valentino}}, \bibinfo {author} {\bibfnamefont {S.}~\bibnamefont
  {Gariazzo}}, \bibinfo {author} {\bibfnamefont {E.}~\bibnamefont {Giusarma}},
  \ and\ \bibinfo {author} {\bibfnamefont {O.}~\bibnamefont {Mena}},\ }\href
  {\doibase 10.1103/PhysRevD.91.123505} {\bibfield  {journal} {\bibinfo
  {journal} {Phys. Rev. D}\ }\textbf {\bibinfo {volume} {91}},\ \bibinfo
  {pages} {123505} (\bibinfo {year} {2015})},\ \Eprint
  {http://arxiv.org/abs/1503.00911} {arXiv:1503.00911 [astro-ph.CO]}
  \BibitemShut {NoStop}%
\bibitem [{\citenamefont {Di~Valentino}\ \emph {et~al.}(2016)\citenamefont
  {Di~Valentino}, \citenamefont {Giusarma}, \citenamefont {Lattanzi},
  \citenamefont {Mena}, \citenamefont {Melchiorri},\ and\ \citenamefont
  {Silk}}]{DiValentino:2015wba}%
  \BibitemOpen
  \bibfield  {author} {\bibinfo {author} {\bibfnamefont {E.}~\bibnamefont
  {Di~Valentino}}, \bibinfo {author} {\bibfnamefont {E.}~\bibnamefont
  {Giusarma}}, \bibinfo {author} {\bibfnamefont {M.}~\bibnamefont {Lattanzi}},
  \bibinfo {author} {\bibfnamefont {O.}~\bibnamefont {Mena}}, \bibinfo {author}
  {\bibfnamefont {A.}~\bibnamefont {Melchiorri}}, \ and\ \bibinfo {author}
  {\bibfnamefont {J.}~\bibnamefont {Silk}},\ }\href {\doibase
  10.1016/j.physletb.2015.11.025} {\bibfield  {journal} {\bibinfo  {journal}
  {Phys. Lett. B}\ }\textbf {\bibinfo {volume} {752}},\ \bibinfo {pages} {182}
  (\bibinfo {year} {2016})},\ \Eprint {http://arxiv.org/abs/1507.08665}
  {arXiv:1507.08665 [astro-ph.CO]} \BibitemShut {NoStop}%
\bibitem [{\citenamefont {Archidiacono}\ \emph {et~al.}(2015)\citenamefont
  {Archidiacono}, \citenamefont {Basse}, \citenamefont {Hamann}, \citenamefont
  {Hannestad}, \citenamefont {Raffelt},\ and\ \citenamefont
  {Wong}}]{Archidiacono:2015mda}%
  \BibitemOpen
  \bibfield  {author} {\bibinfo {author} {\bibfnamefont {M.}~\bibnamefont
  {Archidiacono}}, \bibinfo {author} {\bibfnamefont {T.}~\bibnamefont {Basse}},
  \bibinfo {author} {\bibfnamefont {J.}~\bibnamefont {Hamann}}, \bibinfo
  {author} {\bibfnamefont {S.}~\bibnamefont {Hannestad}}, \bibinfo {author}
  {\bibfnamefont {G.}~\bibnamefont {Raffelt}}, \ and\ \bibinfo {author}
  {\bibfnamefont {Y.~Y.}\ \bibnamefont {Wong}},\ }\href {\doibase
  10.1088/1475-7516/2015/05/050} {\bibfield  {journal} {\bibinfo  {journal}
  {JCAP}\ }\textbf {\bibinfo {volume} {05}},\ \bibinfo {pages} {050} (\bibinfo
  {year} {2015})},\ \Eprint {http://arxiv.org/abs/1502.03325} {arXiv:1502.03325
  [astro-ph.CO]} \BibitemShut {NoStop}%
\bibitem [{\citenamefont {Giar\`e}\ \emph {et~al.}(2020)\citenamefont
  {Giar\`e}, \citenamefont {Di~Valentino}, \citenamefont {Melchiorri},\ and\
  \citenamefont {Mena}}]{Giare:2020vzo}%
  \BibitemOpen
  \bibfield  {author} {\bibinfo {author} {\bibfnamefont {W.}~\bibnamefont
  {Giar\`e}}, \bibinfo {author} {\bibfnamefont {E.}~\bibnamefont
  {Di~Valentino}}, \bibinfo {author} {\bibfnamefont {A.}~\bibnamefont
  {Melchiorri}}, \ and\ \bibinfo {author} {\bibfnamefont {O.}~\bibnamefont
  {Mena}},\ }\href@noop {} {\  (\bibinfo {year} {2020})},\ \Eprint
  {http://arxiv.org/abs/2011.14704} {arXiv:2011.14704 [astro-ph.CO]}
  \BibitemShut {NoStop}%
\bibitem [{\citenamefont {Kim}(1979)}]{Kim:1979if}%
  \BibitemOpen
  \bibfield  {author} {\bibinfo {author} {\bibfnamefont {J.~E.}\ \bibnamefont
  {Kim}},\ }\href {\doibase 10.1103/PhysRevLett.43.103} {\bibfield  {journal}
  {\bibinfo  {journal} {Phys. Rev. Lett.}\ }\textbf {\bibinfo {volume} {43}},\
  \bibinfo {pages} {103} (\bibinfo {year} {1979})}\BibitemShut {NoStop}%
%%CITATION = PRLTA,43,103;%%
\bibitem [{\citenamefont {Shifman}\ \emph {et~al.}(1980)\citenamefont
  {Shifman}, \citenamefont {Vainshtein},\ and\ \citenamefont
  {Zakharov}}]{Shifman:1979if}%
  \BibitemOpen
  \bibfield  {author} {\bibinfo {author} {\bibfnamefont {M.~A.}\ \bibnamefont
  {Shifman}}, \bibinfo {author} {\bibfnamefont {A.~I.}\ \bibnamefont
  {Vainshtein}}, \ and\ \bibinfo {author} {\bibfnamefont {V.~I.}\ \bibnamefont
  {Zakharov}},\ }\href {\doibase 10.1016/0550-3213(80)90209-6} {\bibfield
  {journal} {\bibinfo  {journal} {Nucl. Phys.}\ }\textbf {\bibinfo {volume}
  {B166}},\ \bibinfo {pages} {493} (\bibinfo {year} {1980})}\BibitemShut
  {NoStop}%
%%CITATION = NUPHA,B166,493;%%
\bibitem [{\citenamefont {Kaplan}(1985)}]{Kaplan:1985dv}%
  \BibitemOpen
  \bibfield  {author} {\bibinfo {author} {\bibfnamefont {D.~B.}\ \bibnamefont
  {Kaplan}},\ }\href {\doibase 10.1016/0550-3213(85)90319-0} {\bibfield
  {journal} {\bibinfo  {journal} {Nucl. Phys. B}\ }\textbf {\bibinfo {volume}
  {260}},\ \bibinfo {pages} {215} (\bibinfo {year} {1985})}\BibitemShut
  {NoStop}%
\bibitem [{\citenamefont {Di~Luzio}\ \emph
  {et~al.}(2017{\natexlab{a}})\citenamefont {Di~Luzio}, \citenamefont
  {Mescia},\ and\ \citenamefont {Nardi}}]{DiLuzio:2016sbl}%
  \BibitemOpen
  \bibfield  {author} {\bibinfo {author} {\bibfnamefont {L.}~\bibnamefont
  {Di~Luzio}}, \bibinfo {author} {\bibfnamefont {F.}~\bibnamefont {Mescia}}, \
  and\ \bibinfo {author} {\bibfnamefont {E.}~\bibnamefont {Nardi}},\ }\href
  {\doibase 10.1103/PhysRevLett.118.031801} {\bibfield  {journal} {\bibinfo
  {journal} {Phys. Rev. Lett.}\ }\textbf {\bibinfo {volume} {118}},\ \bibinfo
  {pages} {031801} (\bibinfo {year} {2017}{\natexlab{a}})},\ \Eprint
  {http://arxiv.org/abs/1610.07593} {arXiv:1610.07593 [hep-ph]} \BibitemShut
  {NoStop}%
\bibitem [{\citenamefont {Di~Luzio}\ \emph
  {et~al.}(2017{\natexlab{b}})\citenamefont {Di~Luzio}, \citenamefont
  {Mescia},\ and\ \citenamefont {Nardi}}]{DiLuzio:2017pfr}%
  \BibitemOpen
  \bibfield  {author} {\bibinfo {author} {\bibfnamefont {L.}~\bibnamefont
  {Di~Luzio}}, \bibinfo {author} {\bibfnamefont {F.}~\bibnamefont {Mescia}}, \
  and\ \bibinfo {author} {\bibfnamefont {E.}~\bibnamefont {Nardi}},\ }\href
  {\doibase 10.1103/PhysRevD.96.075003} {\bibfield  {journal} {\bibinfo
  {journal} {Phys. Rev. D}\ }\textbf {\bibinfo {volume} {96}},\ \bibinfo
  {pages} {075003} (\bibinfo {year} {2017}{\natexlab{b}})},\ \Eprint
  {http://arxiv.org/abs/1705.05370} {arXiv:1705.05370 [hep-ph]} \BibitemShut
  {NoStop}%
\bibitem [{\citenamefont {Raffelt}(1990)}]{Raffelt:1990yz}%
  \BibitemOpen
  \bibfield  {author} {\bibinfo {author} {\bibfnamefont {G.~G.}\ \bibnamefont
  {Raffelt}},\ }\href {\doibase 10.1016/0370-1573(90)90054-6} {\bibfield
  {journal} {\bibinfo  {journal} {Phys. Rept.}\ }\textbf {\bibinfo {volume}
  {198}},\ \bibinfo {pages} {1} (\bibinfo {year} {1990})}\BibitemShut {NoStop}%
\bibitem [{\citenamefont {Chang}\ \emph {et~al.}(2018)\citenamefont {Chang},
  \citenamefont {Essig},\ and\ \citenamefont {McDermott}}]{Chang:2018rso}%
  \BibitemOpen
  \bibfield  {author} {\bibinfo {author} {\bibfnamefont {J.~H.}\ \bibnamefont
  {Chang}}, \bibinfo {author} {\bibfnamefont {R.}~\bibnamefont {Essig}}, \ and\
  \bibinfo {author} {\bibfnamefont {S.~D.}\ \bibnamefont {McDermott}},\ }\href
  {\doibase 10.1007/JHEP09(2018)051} {\bibfield  {journal} {\bibinfo  {journal}
  {JHEP}\ }\textbf {\bibinfo {volume} {09}},\ \bibinfo {pages} {051} (\bibinfo
  {year} {2018})},\ \Eprint {http://arxiv.org/abs/1803.00993} {arXiv:1803.00993
  [hep-ph]} \BibitemShut {NoStop}%
\bibitem [{\citenamefont {Carenza}\ \emph {et~al.}(2019)\citenamefont
  {Carenza}, \citenamefont {Fischer}, \citenamefont {Giannotti}, \citenamefont
  {Guo}, \citenamefont {Mart\'\i{}nez-Pinedo},\ and\ \citenamefont
  {Mirizzi}}]{Carenza:2019pxu}%
  \BibitemOpen
  \bibfield  {author} {\bibinfo {author} {\bibfnamefont {P.}~\bibnamefont
  {Carenza}}, \bibinfo {author} {\bibfnamefont {T.}~\bibnamefont {Fischer}},
  \bibinfo {author} {\bibfnamefont {M.}~\bibnamefont {Giannotti}}, \bibinfo
  {author} {\bibfnamefont {G.}~\bibnamefont {Guo}}, \bibinfo {author}
  {\bibfnamefont {G.}~\bibnamefont {Mart\'\i{}nez-Pinedo}}, \ and\ \bibinfo
  {author} {\bibfnamefont {A.}~\bibnamefont {Mirizzi}},\ }\href {\doibase
  10.1088/1475-7516/2019/10/016} {\bibfield  {journal} {\bibinfo  {journal}
  {JCAP}\ }\textbf {\bibinfo {volume} {10}},\ \bibinfo {pages} {016} (\bibinfo
  {year} {2019})},\ \bibinfo {note} {[Erratum: JCAP 05, E01 (2020)]},\ \Eprint
  {http://arxiv.org/abs/1906.11844} {arXiv:1906.11844 [hep-ph]} \BibitemShut
  {NoStop}%
\bibitem [{\citenamefont {Bar}\ \emph {et~al.}(2020)\citenamefont {Bar},
  \citenamefont {Blum},\ and\ \citenamefont {D'Amico}}]{Bar:2019ifz}%
  \BibitemOpen
  \bibfield  {author} {\bibinfo {author} {\bibfnamefont {N.}~\bibnamefont
  {Bar}}, \bibinfo {author} {\bibfnamefont {K.}~\bibnamefont {Blum}}, \ and\
  \bibinfo {author} {\bibfnamefont {G.}~\bibnamefont {D'Amico}},\ }\href
  {\doibase 10.1103/PhysRevD.101.123025} {\bibfield  {journal} {\bibinfo
  {journal} {Phys. Rev. D}\ }\textbf {\bibinfo {volume} {101}},\ \bibinfo
  {pages} {123025} (\bibinfo {year} {2020})},\ \Eprint
  {http://arxiv.org/abs/1907.05020} {arXiv:1907.05020 [hep-ph]} \BibitemShut
  {NoStop}%
\bibitem [{\citenamefont {Bazavov}\ \emph {et~al.}(2012)\citenamefont {Bazavov}
  \emph {et~al.}}]{Bazavov:2011nk}%
  \BibitemOpen
  \bibfield  {author} {\bibinfo {author} {\bibfnamefont {A.}~\bibnamefont
  {Bazavov}} \emph {et~al.},\ }\href {\doibase 10.1103/PhysRevD.85.054503}
  {\bibfield  {journal} {\bibinfo  {journal} {Phys. Rev. D}\ }\textbf {\bibinfo
  {volume} {85}},\ \bibinfo {pages} {054503} (\bibinfo {year} {2012})},\
  \Eprint {http://arxiv.org/abs/1111.1710} {arXiv:1111.1710 [hep-lat]}
  \BibitemShut {NoStop}%
\bibitem [{\citenamefont {Aoki}\ \emph {et~al.}(2006)\citenamefont {Aoki},
  \citenamefont {Fodor}, \citenamefont {Katz},\ and\ \citenamefont
  {Szabo}}]{Aoki:2006br}%
  \BibitemOpen
  \bibfield  {author} {\bibinfo {author} {\bibfnamefont {Y.}~\bibnamefont
  {Aoki}}, \bibinfo {author} {\bibfnamefont {Z.}~\bibnamefont {Fodor}},
  \bibinfo {author} {\bibfnamefont {S.~D.}\ \bibnamefont {Katz}}, \ and\
  \bibinfo {author} {\bibfnamefont {K.~K.}\ \bibnamefont {Szabo}},\ }\href
  {\doibase 10.1016/j.physletb.2006.10.021} {\bibfield  {journal} {\bibinfo
  {journal} {Phys. Lett. B}\ }\textbf {\bibinfo {volume} {643}},\ \bibinfo
  {pages} {46} (\bibinfo {year} {2006})},\ \Eprint
  {http://arxiv.org/abs/hep-lat/0609068} {arXiv:hep-lat/0609068} \BibitemShut
  {NoStop}%
\bibitem [{\citenamefont {Borsanyi}\ \emph {et~al.}(2010)\citenamefont
  {Borsanyi}, \citenamefont {Fodor}, \citenamefont {Hoelbling}, \citenamefont
  {Katz}, \citenamefont {Krieg}, \citenamefont {Ratti},\ and\ \citenamefont
  {Szabo}}]{Borsanyi:2010bp}%
  \BibitemOpen
  \bibfield  {author} {\bibinfo {author} {\bibfnamefont {S.}~\bibnamefont
  {Borsanyi}}, \bibinfo {author} {\bibfnamefont {Z.}~\bibnamefont {Fodor}},
  \bibinfo {author} {\bibfnamefont {C.}~\bibnamefont {Hoelbling}}, \bibinfo
  {author} {\bibfnamefont {S.~D.}\ \bibnamefont {Katz}}, \bibinfo {author}
  {\bibfnamefont {S.}~\bibnamefont {Krieg}}, \bibinfo {author} {\bibfnamefont
  {C.}~\bibnamefont {Ratti}}, \ and\ \bibinfo {author} {\bibfnamefont {K.~K.}\
  \bibnamefont {Szabo}} (\bibinfo {collaboration} {Wuppertal-Budapest}),\
  }\href {\doibase 10.1007/JHEP09(2010)073} {\bibfield  {journal} {\bibinfo
  {journal} {JHEP}\ }\textbf {\bibinfo {volume} {09}},\ \bibinfo {pages} {073}
  (\bibinfo {year} {2010})},\ \Eprint {http://arxiv.org/abs/1005.3508}
  {arXiv:1005.3508 [hep-lat]} \BibitemShut {NoStop}%
\bibitem [{\citenamefont {Spalinski}(1988)}]{Spalinski:1988az}%
  \BibitemOpen
  \bibfield  {author} {\bibinfo {author} {\bibfnamefont {M.}~\bibnamefont
  {Spalinski}},\ }\href {\doibase 10.1007/BF01412582} {\bibfield  {journal}
  {\bibinfo  {journal} {Z. Phys. C}\ }\textbf {\bibinfo {volume} {41}},\
  \bibinfo {pages} {87} (\bibinfo {year} {1988})}\BibitemShut {NoStop}%
\bibitem [{\citenamefont {Grilli~di Cortona}\ \emph {et~al.}(2016)\citenamefont
  {Grilli~di Cortona}, \citenamefont {Hardy}, \citenamefont {Pardo~Vega},\ and\
  \citenamefont {Villadoro}}]{diCortona:2015ldu}%
  \BibitemOpen
  \bibfield  {author} {\bibinfo {author} {\bibfnamefont {G.}~\bibnamefont
  {Grilli~di Cortona}}, \bibinfo {author} {\bibfnamefont {E.}~\bibnamefont
  {Hardy}}, \bibinfo {author} {\bibfnamefont {J.}~\bibnamefont {Pardo~Vega}}, \
  and\ \bibinfo {author} {\bibfnamefont {G.}~\bibnamefont {Villadoro}},\ }\href
  {\doibase 10.1007/JHEP01(2016)034} {\bibfield  {journal} {\bibinfo  {journal}
  {JHEP}\ }\textbf {\bibinfo {volume} {01}},\ \bibinfo {pages} {034} (\bibinfo
  {year} {2016})},\ \Eprint {http://arxiv.org/abs/1511.02867} {arXiv:1511.02867
  [hep-ph]} \BibitemShut {NoStop}%
\bibitem [{\citenamefont {Di~Vecchia}\ and\ \citenamefont
  {Veneziano}(1980)}]{DiVecchia:1980yfw}%
  \BibitemOpen
  \bibfield  {author} {\bibinfo {author} {\bibfnamefont {P.}~\bibnamefont
  {Di~Vecchia}}\ and\ \bibinfo {author} {\bibfnamefont {G.}~\bibnamefont
  {Veneziano}},\ }\href {\doibase 10.1016/0550-3213(80)90370-3} {\bibfield
  {journal} {\bibinfo  {journal} {Nucl. Phys. B}\ }\textbf {\bibinfo {volume}
  {171}},\ \bibinfo {pages} {253} (\bibinfo {year} {1980})}\BibitemShut
  {NoStop}%
\bibitem [{\citenamefont {Georgi}\ \emph {et~al.}(1986)\citenamefont {Georgi},
  \citenamefont {Kaplan},\ and\ \citenamefont {Randall}}]{Georgi:1986df}%
  \BibitemOpen
  \bibfield  {author} {\bibinfo {author} {\bibfnamefont {H.}~\bibnamefont
  {Georgi}}, \bibinfo {author} {\bibfnamefont {D.~B.}\ \bibnamefont {Kaplan}},
  \ and\ \bibinfo {author} {\bibfnamefont {L.}~\bibnamefont {Randall}},\ }\href
  {\doibase 10.1016/0370-2693(86)90688-X} {\bibfield  {journal} {\bibinfo
  {journal} {Phys. Lett. B}\ }\textbf {\bibinfo {volume} {169}},\ \bibinfo
  {pages} {73} (\bibinfo {year} {1986})}\BibitemShut {NoStop}%
\bibitem [{\citenamefont {Di~Luzio}\ \emph {et~al.}(2020)\citenamefont
  {Di~Luzio}, \citenamefont {Giannotti}, \citenamefont {Nardi},\ and\
  \citenamefont {Visinelli}}]{DiLuzio:2020wdo}%
  \BibitemOpen
  \bibfield  {author} {\bibinfo {author} {\bibfnamefont {L.}~\bibnamefont
  {Di~Luzio}}, \bibinfo {author} {\bibfnamefont {M.}~\bibnamefont {Giannotti}},
  \bibinfo {author} {\bibfnamefont {E.}~\bibnamefont {Nardi}}, \ and\ \bibinfo
  {author} {\bibfnamefont {L.}~\bibnamefont {Visinelli}},\ }\href {\doibase
  10.1016/j.physrep.2020.06.002} {\bibfield  {journal} {\bibinfo  {journal}
  {Phys. Rept.}\ }\textbf {\bibinfo {volume} {870}},\ \bibinfo {pages} {1}
  (\bibinfo {year} {2020})},\ \Eprint {http://arxiv.org/abs/2003.01100}
  {arXiv:2003.01100 [hep-ph]} \BibitemShut {NoStop}%
\bibitem [{\citenamefont {Zhitnitsky}(1980)}]{Zhitnitsky:1980tq}%
  \BibitemOpen
  \bibfield  {author} {\bibinfo {author} {\bibfnamefont {A.~R.}\ \bibnamefont
  {Zhitnitsky}},\ }\href@noop {} {\bibfield  {journal} {\bibinfo  {journal}
  {Sov. J. Nucl. Phys.}\ }\textbf {\bibinfo {volume} {31}},\ \bibinfo {pages}
  {260} (\bibinfo {year} {1980})},\ \bibinfo {note} {[Yad.
  Fiz.31,497(1980)]}\BibitemShut {NoStop}%
%%CITATION = SJNCA,31,260;%%
\bibitem [{\citenamefont {Dine}\ \emph {et~al.}(1981)\citenamefont {Dine},
  \citenamefont {Fischler},\ and\ \citenamefont {Srednicki}}]{Dine:1981rt}%
  \BibitemOpen
  \bibfield  {author} {\bibinfo {author} {\bibfnamefont {M.}~\bibnamefont
  {Dine}}, \bibinfo {author} {\bibfnamefont {W.}~\bibnamefont {Fischler}}, \
  and\ \bibinfo {author} {\bibfnamefont {M.}~\bibnamefont {Srednicki}},\ }\href
  {\doibase 10.1016/0370-2693(81)90590-6} {\bibfield  {journal} {\bibinfo
  {journal} {Phys. Lett.}\ }\textbf {\bibinfo {volume} {B104}},\ \bibinfo
  {pages} {199} (\bibinfo {year} {1981})}\BibitemShut {NoStop}%
%%CITATION = PHLTA,B104,199;%%
\bibitem [{\citenamefont {Gasser}\ and\ \citenamefont
  {Leutwyler}(1984)}]{Gasser:1983yg}%
  \BibitemOpen
  \bibfield  {author} {\bibinfo {author} {\bibfnamefont {J.}~\bibnamefont
  {Gasser}}\ and\ \bibinfo {author} {\bibfnamefont {H.}~\bibnamefont
  {Leutwyler}},\ }\href {\doibase 10.1016/0003-4916(84)90242-2} {\bibfield
  {journal} {\bibinfo  {journal} {Annals Phys.}\ }\textbf {\bibinfo {volume}
  {158}},\ \bibinfo {pages} {142} (\bibinfo {year} {1984})}\BibitemShut
  {NoStop}%
\bibitem [{\citenamefont {Scherer}(2003)}]{Scherer:2002tk}%
  \BibitemOpen
  \bibfield  {author} {\bibinfo {author} {\bibfnamefont {S.}~\bibnamefont
  {Scherer}},\ }\href@noop {} {\bibfield  {journal} {\bibinfo  {journal} {Adv.
  Nucl. Phys.}\ }\textbf {\bibinfo {volume} {27}},\ \bibinfo {pages} {277}
  (\bibinfo {year} {2003})},\ \Eprint {http://arxiv.org/abs/hep-ph/0210398}
  {arXiv:hep-ph/0210398} \BibitemShut {NoStop}%
\bibitem [{\citenamefont {Wess}\ and\ \citenamefont
  {Zumino}(1971)}]{Wess:1971yu}%
  \BibitemOpen
  \bibfield  {author} {\bibinfo {author} {\bibfnamefont {J.}~\bibnamefont
  {Wess}}\ and\ \bibinfo {author} {\bibfnamefont {B.}~\bibnamefont {Zumino}},\
  }\href {\doibase 10.1016/0370-2693(71)90582-X} {\bibfield  {journal}
  {\bibinfo  {journal} {Phys. Lett. B}\ }\textbf {\bibinfo {volume} {37}},\
  \bibinfo {pages} {95} (\bibinfo {year} {1971})}\BibitemShut {NoStop}%
\bibitem [{\citenamefont {Witten}(1983)}]{Witten:1983tw}%
  \BibitemOpen
  \bibfield  {author} {\bibinfo {author} {\bibfnamefont {E.}~\bibnamefont
  {Witten}},\ }\href {\doibase 10.1016/0550-3213(83)90063-9} {\bibfield
  {journal} {\bibinfo  {journal} {Nucl. Phys. B}\ }\textbf {\bibinfo {volume}
  {223}},\ \bibinfo {pages} {422} (\bibinfo {year} {1983})}\BibitemShut
  {NoStop}%
\bibitem [{\citenamefont {Lehmann}\ \emph {et~al.}(1955)\citenamefont
  {Lehmann}, \citenamefont {Symanzik},\ and\ \citenamefont
  {Zimmermann}}]{Lehmann:1954rq}%
  \BibitemOpen
  \bibfield  {author} {\bibinfo {author} {\bibfnamefont {H.}~\bibnamefont
  {Lehmann}}, \bibinfo {author} {\bibfnamefont {K.}~\bibnamefont {Symanzik}}, \
  and\ \bibinfo {author} {\bibfnamefont {W.}~\bibnamefont {Zimmermann}},\
  }\href {\doibase 10.1007/BF02731765} {\bibfield  {journal} {\bibinfo
  {journal} {Nuovo Cim.}\ }\textbf {\bibinfo {volume} {1}},\ \bibinfo {pages}
  {205} (\bibinfo {year} {1955})}\BibitemShut {NoStop}%
\bibitem [{\citenamefont {Ferreira}\ \emph {et~al.}(2020)\citenamefont
  {Ferreira}, \citenamefont {Notari},\ and\ \citenamefont
  {Rompineve}}]{Ferreira:2020bpb}%
  \BibitemOpen
  \bibfield  {author} {\bibinfo {author} {\bibfnamefont {R.~Z.}\ \bibnamefont
  {Ferreira}}, \bibinfo {author} {\bibfnamefont {A.}~\bibnamefont {Notari}}, \
  and\ \bibinfo {author} {\bibfnamefont {F.}~\bibnamefont {Rompineve}},\
  }\href@noop {} {\  (\bibinfo {year} {2020})},\ \Eprint
  {http://arxiv.org/abs/2012.06566} {arXiv:2012.06566 [hep-ph]} \BibitemShut
  {NoStop}%
\bibitem [{\citenamefont {Gasser}\ and\ \citenamefont
  {Leutwyler}(1987{\natexlab{a}})}]{Gasser:1986vb}%
  \BibitemOpen
  \bibfield  {author} {\bibinfo {author} {\bibfnamefont {J.}~\bibnamefont
  {Gasser}}\ and\ \bibinfo {author} {\bibfnamefont {H.}~\bibnamefont
  {Leutwyler}},\ }\href {\doibase 10.1016/0370-2693(87)90492-8} {\bibfield
  {journal} {\bibinfo  {journal} {Phys. Lett. B}\ }\textbf {\bibinfo {volume}
  {184}},\ \bibinfo {pages} {83} (\bibinfo {year}
  {1987}{\natexlab{a}})}\BibitemShut {NoStop}%
\bibitem [{\citenamefont {Gasser}\ and\ \citenamefont
  {Leutwyler}(1987{\natexlab{b}})}]{Gasser:1987ah}%
  \BibitemOpen
  \bibfield  {author} {\bibinfo {author} {\bibfnamefont {J.}~\bibnamefont
  {Gasser}}\ and\ \bibinfo {author} {\bibfnamefont {H.}~\bibnamefont
  {Leutwyler}},\ }\href {\doibase 10.1016/0370-2693(87)91652-2} {\bibfield
  {journal} {\bibinfo  {journal} {Phys. Lett. B}\ }\textbf {\bibinfo {volume}
  {188}},\ \bibinfo {pages} {477} (\bibinfo {year}
  {1987}{\natexlab{b}})}\BibitemShut {NoStop}%
\bibitem [{\citenamefont {Gerber}\ and\ \citenamefont
  {Leutwyler}(1989)}]{Gerber:1988tt}%
  \BibitemOpen
  \bibfield  {author} {\bibinfo {author} {\bibfnamefont {P.}~\bibnamefont
  {Gerber}}\ and\ \bibinfo {author} {\bibfnamefont {H.}~\bibnamefont
  {Leutwyler}},\ }\href {\doibase 10.1016/0550-3213(89)90349-0} {\bibfield
  {journal} {\bibinfo  {journal} {Nucl. Phys. B}\ }\textbf {\bibinfo {volume}
  {321}},\ \bibinfo {pages} {387} (\bibinfo {year} {1989})}\BibitemShut
  {NoStop}%
\bibitem [{\citenamefont {Hannestad}\ and\ \citenamefont
  {Madsen}(1995)}]{Hannestad:1995rs}%
  \BibitemOpen
  \bibfield  {author} {\bibinfo {author} {\bibfnamefont {S.}~\bibnamefont
  {Hannestad}}\ and\ \bibinfo {author} {\bibfnamefont {J.}~\bibnamefont
  {Madsen}},\ }\href {\doibase 10.1103/PhysRevD.52.1764} {\bibfield  {journal}
  {\bibinfo  {journal} {Phys. Rev. D}\ }\textbf {\bibinfo {volume} {52}},\
  \bibinfo {pages} {1764} (\bibinfo {year} {1995})},\ \Eprint
  {http://arxiv.org/abs/astro-ph/9506015} {arXiv:astro-ph/9506015} \BibitemShut
  {NoStop}%
\bibitem [{\citenamefont {Colangelo}\ \emph {et~al.}(2001)\citenamefont
  {Colangelo}, \citenamefont {Gasser},\ and\ \citenamefont
  {Leutwyler}}]{Colangelo:2001df}%
  \BibitemOpen
  \bibfield  {author} {\bibinfo {author} {\bibfnamefont {G.}~\bibnamefont
  {Colangelo}}, \bibinfo {author} {\bibfnamefont {J.}~\bibnamefont {Gasser}}, \
  and\ \bibinfo {author} {\bibfnamefont {H.}~\bibnamefont {Leutwyler}},\ }\href
  {\doibase 10.1016/S0550-3213(01)00147-X} {\bibfield  {journal} {\bibinfo
  {journal} {Nucl. Phys. B}\ }\textbf {\bibinfo {volume} {603}},\ \bibinfo
  {pages} {125} (\bibinfo {year} {2001})},\ \Eprint
  {http://arxiv.org/abs/hep-ph/0103088} {arXiv:hep-ph/0103088} \BibitemShut
  {NoStop}%
\bibitem [{\citenamefont {Aoki}\ \emph {et~al.}(2020)\citenamefont {Aoki} \emph
  {et~al.}}]{Aoki:2019cca}%
  \BibitemOpen
  \bibfield  {author} {\bibinfo {author} {\bibfnamefont {S.}~\bibnamefont
  {Aoki}} \emph {et~al.} (\bibinfo {collaboration} {Flavour Lattice Averaging
  Group}),\ }\href {\doibase 10.1140/epjc/s10052-019-7354-7} {\bibfield
  {journal} {\bibinfo  {journal} {Eur. Phys. J. C}\ }\textbf {\bibinfo {volume}
  {80}},\ \bibinfo {pages} {113} (\bibinfo {year} {2020})},\ \Eprint
  {http://arxiv.org/abs/1902.08191} {arXiv:1902.08191 [hep-lat]} \BibitemShut
  {NoStop}%
\bibitem [{\citenamefont {Darm\'e}\ \emph {et~al.}(2020)\citenamefont
  {Darm\'e}, \citenamefont {Di~Luzio}, \citenamefont {Giannotti},\ and\
  \citenamefont {Nardi}}]{Darme:2020gyx}%
  \BibitemOpen
  \bibfield  {author} {\bibinfo {author} {\bibfnamefont {L.}~\bibnamefont
  {Darm\'e}}, \bibinfo {author} {\bibfnamefont {L.}~\bibnamefont {Di~Luzio}},
  \bibinfo {author} {\bibfnamefont {M.}~\bibnamefont {Giannotti}}, \ and\
  \bibinfo {author} {\bibfnamefont {E.}~\bibnamefont {Nardi}},\ }\href@noop {}
  {\  (\bibinfo {year} {2020})},\ \Eprint {http://arxiv.org/abs/2010.15846}
  {arXiv:2010.15846 [hep-ph]} \BibitemShut {NoStop}%
\bibitem [{\citenamefont {Weinberg}(1966)}]{Weinberg:1966kf}%
  \BibitemOpen
  \bibfield  {author} {\bibinfo {author} {\bibfnamefont {S.}~\bibnamefont
  {Weinberg}},\ }\href {\doibase 10.1103/PhysRevLett.17.616} {\bibfield
  {journal} {\bibinfo  {journal} {Phys. Rev. Lett.}\ }\textbf {\bibinfo
  {volume} {17}},\ \bibinfo {pages} {616} (\bibinfo {year} {1966})}\BibitemShut
  {NoStop}%
\bibitem [{\citenamefont {Aydemir}\ \emph {et~al.}(2012)\citenamefont
  {Aydemir}, \citenamefont {Anber},\ and\ \citenamefont
  {Donoghue}}]{Aydemir:2012nz}%
  \BibitemOpen
  \bibfield  {author} {\bibinfo {author} {\bibfnamefont {U.}~\bibnamefont
  {Aydemir}}, \bibinfo {author} {\bibfnamefont {M.~M.}\ \bibnamefont {Anber}},
  \ and\ \bibinfo {author} {\bibfnamefont {J.~F.}\ \bibnamefont {Donoghue}},\
  }\href {\doibase 10.1103/PhysRevD.86.014025} {\bibfield  {journal} {\bibinfo
  {journal} {Phys. Rev. D}\ }\textbf {\bibinfo {volume} {86}},\ \bibinfo
  {pages} {014025} (\bibinfo {year} {2012})},\ \Eprint
  {http://arxiv.org/abs/1203.5153} {arXiv:1203.5153 [hep-ph]} \BibitemShut
  {NoStop}%
\bibitem [{\citenamefont {Blum}\ \emph {et~al.}(2015)\citenamefont {Blum} \emph
  {et~al.}}]{Blum:2015ywa}%
  \BibitemOpen
  \bibfield  {author} {\bibinfo {author} {\bibfnamefont {T.}~\bibnamefont
  {Blum}} \emph {et~al.},\ }\href {\doibase 10.1103/PhysRevD.91.074502}
  {\bibfield  {journal} {\bibinfo  {journal} {Phys. Rev. D}\ }\textbf {\bibinfo
  {volume} {91}},\ \bibinfo {pages} {074502} (\bibinfo {year} {2015})},\
  \Eprint {http://arxiv.org/abs/1502.00263} {arXiv:1502.00263 [hep-lat]}
  \BibitemShut {NoStop}%
\bibitem [{\citenamefont {Abbott}\ \emph {et~al.}(2020)\citenamefont {Abbott}
  \emph {et~al.}}]{Abbott:2020hxn}%
  \BibitemOpen
  \bibfield  {author} {\bibinfo {author} {\bibfnamefont {R.}~\bibnamefont
  {Abbott}} \emph {et~al.} (\bibinfo {collaboration} {RBC, UKQCD}),\ }\href
  {\doibase 10.1103/PhysRevD.102.054509} {\bibfield  {journal} {\bibinfo
  {journal} {Phys. Rev. D}\ }\textbf {\bibinfo {volume} {102}},\ \bibinfo
  {pages} {054509} (\bibinfo {year} {2020})},\ \Eprint
  {http://arxiv.org/abs/2004.09440} {arXiv:2004.09440 [hep-lat]} \BibitemShut
  {NoStop}%
\bibitem [{\citenamefont {Luscher}(1991)}]{Luscher:1990ux}%
  \BibitemOpen
  \bibfield  {author} {\bibinfo {author} {\bibfnamefont {M.}~\bibnamefont
  {Luscher}},\ }\href {\doibase 10.1016/0550-3213(91)90366-6} {\bibfield
  {journal} {\bibinfo  {journal} {Nucl. Phys. B}\ }\textbf {\bibinfo {volume}
  {354}},\ \bibinfo {pages} {531} (\bibinfo {year} {1991})}\BibitemShut
  {NoStop}%
\bibitem [{\citenamefont {Rummukainen}\ and\ \citenamefont
  {Gottlieb}(1995)}]{Rummukainen:1995vs}%
  \BibitemOpen
  \bibfield  {author} {\bibinfo {author} {\bibfnamefont {K.}~\bibnamefont
  {Rummukainen}}\ and\ \bibinfo {author} {\bibfnamefont {S.~A.}\ \bibnamefont
  {Gottlieb}},\ }\href {\doibase 10.1016/0550-3213(95)00313-H} {\bibfield
  {journal} {\bibinfo  {journal} {Nucl. Phys. B}\ }\textbf {\bibinfo {volume}
  {450}},\ \bibinfo {pages} {397} (\bibinfo {year} {1995})},\ \Eprint
  {http://arxiv.org/abs/hep-lat/9503028} {arXiv:hep-lat/9503028} \BibitemShut
  {NoStop}%
\bibitem [{\citenamefont {Kim}\ \emph {et~al.}(2005)\citenamefont {Kim},
  \citenamefont {Sachrajda},\ and\ \citenamefont {Sharpe}}]{Kim:2005gf}%
  \BibitemOpen
  \bibfield  {author} {\bibinfo {author} {\bibfnamefont {C.~h.}\ \bibnamefont
  {Kim}}, \bibinfo {author} {\bibfnamefont {C.~T.}\ \bibnamefont {Sachrajda}},
  \ and\ \bibinfo {author} {\bibfnamefont {S.~R.}\ \bibnamefont {Sharpe}},\
  }\href {\doibase 10.1016/j.nuclphysb.2005.08.029} {\bibfield  {journal}
  {\bibinfo  {journal} {Nucl. Phys. B}\ }\textbf {\bibinfo {volume} {727}},\
  \bibinfo {pages} {218} (\bibinfo {year} {2005})},\ \Eprint
  {http://arxiv.org/abs/hep-lat/0507006} {arXiv:hep-lat/0507006} \BibitemShut
  {NoStop}%
\bibitem [{\citenamefont {Hansen}\ and\ \citenamefont
  {Sharpe}(2012)}]{Hansen:2012tf}%
  \BibitemOpen
  \bibfield  {author} {\bibinfo {author} {\bibfnamefont {M.~T.}\ \bibnamefont
  {Hansen}}\ and\ \bibinfo {author} {\bibfnamefont {S.~R.}\ \bibnamefont
  {Sharpe}},\ }\href {\doibase 10.1103/PhysRevD.86.016007} {\bibfield
  {journal} {\bibinfo  {journal} {Phys. Rev. D}\ }\textbf {\bibinfo {volume}
  {86}},\ \bibinfo {pages} {016007} (\bibinfo {year} {2012})},\ \Eprint
  {http://arxiv.org/abs/1204.0826} {arXiv:1204.0826 [hep-lat]} \BibitemShut
  {NoStop}%
\bibitem [{\citenamefont {Briceno}\ \emph {et~al.}(2017)\citenamefont
  {Briceno}, \citenamefont {Dudek}, \citenamefont {Edwards},\ and\
  \citenamefont {Wilson}}]{Briceno:2016mjc}%
  \BibitemOpen
  \bibfield  {author} {\bibinfo {author} {\bibfnamefont {R.~A.}\ \bibnamefont
  {Briceno}}, \bibinfo {author} {\bibfnamefont {J.~J.}\ \bibnamefont {Dudek}},
  \bibinfo {author} {\bibfnamefont {R.~G.}\ \bibnamefont {Edwards}}, \ and\
  \bibinfo {author} {\bibfnamefont {D.~J.}\ \bibnamefont {Wilson}},\ }\href
  {\doibase 10.1103/PhysRevLett.118.022002} {\bibfield  {journal} {\bibinfo
  {journal} {Phys. Rev. Lett.}\ }\textbf {\bibinfo {volume} {118}},\ \bibinfo
  {pages} {022002} (\bibinfo {year} {2017})},\ \Eprint
  {http://arxiv.org/abs/1607.05900} {arXiv:1607.05900 [hep-ph]} \BibitemShut
  {NoStop}%
\bibitem [{\citenamefont {Alloul}\ \emph {et~al.}(2014)\citenamefont {Alloul},
  \citenamefont {Christensen}, \citenamefont {Degrande}, \citenamefont {Duhr},\
  and\ \citenamefont {Fuks}}]{FeynRules1}%
  \BibitemOpen
  \bibfield  {author} {\bibinfo {author} {\bibfnamefont {A.}~\bibnamefont
  {Alloul}}, \bibinfo {author} {\bibfnamefont {N.~D.}\ \bibnamefont
  {Christensen}}, \bibinfo {author} {\bibfnamefont {C.}~\bibnamefont
  {Degrande}}, \bibinfo {author} {\bibfnamefont {C.}~\bibnamefont {Duhr}}, \
  and\ \bibinfo {author} {\bibfnamefont {B.}~\bibnamefont {Fuks}},\ }\href
  {\doibase 10.1016/j.cpc.2014.04.012} {\bibfield  {journal} {\bibinfo
  {journal} {Comput. Phys. Commun.}\ }\textbf {\bibinfo {volume} {185}},\
  \bibinfo {pages} {2250} (\bibinfo {year} {2014})},\ \Eprint
  {http://arxiv.org/abs/1310.1921} {arXiv:1310.1921 [hep-ph]} \BibitemShut
  {NoStop}%
\bibitem [{\citenamefont {Christensen}\ and\ \citenamefont
  {Duhr}(2009)}]{feynRules2}%
  \BibitemOpen
  \bibfield  {author} {\bibinfo {author} {\bibfnamefont {N.~D.}\ \bibnamefont
  {Christensen}}\ and\ \bibinfo {author} {\bibfnamefont {C.}~\bibnamefont
  {Duhr}},\ }\href {\doibase 10.1016/j.cpc.2009.02.018} {\bibfield  {journal}
  {\bibinfo  {journal} {Comput. Phys. Commun.}\ }\textbf {\bibinfo {volume}
  {180}},\ \bibinfo {pages} {1614} (\bibinfo {year} {2009})},\ \Eprint
  {http://arxiv.org/abs/0806.4194} {arXiv:0806.4194 [hep-ph]} \BibitemShut
  {NoStop}%
\bibitem [{\citenamefont {Hahn}(2001)}]{feynArts}%
  \BibitemOpen
  \bibfield  {author} {\bibinfo {author} {\bibfnamefont {T.}~\bibnamefont
  {Hahn}},\ }\href {\doibase 10.1016/S0010-4655(01)00290-9} {\bibfield
  {journal} {\bibinfo  {journal} {Comput. Phys. Commun.}\ }\textbf {\bibinfo
  {volume} {140}},\ \bibinfo {pages} {418} (\bibinfo {year} {2001})},\ \Eprint
  {http://arxiv.org/abs/hep-ph/0012260} {arXiv:hep-ph/0012260} \BibitemShut
  {NoStop}%
\bibitem [{\citenamefont {Shtabovenko}\ \emph {et~al.}(2020)\citenamefont
  {Shtabovenko}, \citenamefont {Mertig},\ and\ \citenamefont
  {Orellana}}]{feynCalc1}%
  \BibitemOpen
  \bibfield  {author} {\bibinfo {author} {\bibfnamefont {V.}~\bibnamefont
  {Shtabovenko}}, \bibinfo {author} {\bibfnamefont {R.}~\bibnamefont {Mertig}},
  \ and\ \bibinfo {author} {\bibfnamefont {F.}~\bibnamefont {Orellana}},\
  }\href {\doibase 10.1016/j.cpc.2020.107478} {\bibfield  {journal} {\bibinfo
  {journal} {Comput. Phys. Commun.}\ }\textbf {\bibinfo {volume} {256}},\
  \bibinfo {pages} {107478} (\bibinfo {year} {2020})},\ \Eprint
  {http://arxiv.org/abs/2001.04407} {arXiv:2001.04407 [hep-ph]} \BibitemShut
  {NoStop}%
\bibitem [{\citenamefont {Shtabovenko}\ \emph {et~al.}(2016)\citenamefont
  {Shtabovenko}, \citenamefont {Mertig},\ and\ \citenamefont
  {Orellana}}]{feynCalc2}%
  \BibitemOpen
  \bibfield  {author} {\bibinfo {author} {\bibfnamefont {V.}~\bibnamefont
  {Shtabovenko}}, \bibinfo {author} {\bibfnamefont {R.}~\bibnamefont {Mertig}},
  \ and\ \bibinfo {author} {\bibfnamefont {F.}~\bibnamefont {Orellana}},\
  }\href {\doibase 10.1016/j.cpc.2016.06.008} {\bibfield  {journal} {\bibinfo
  {journal} {Comput. Phys. Commun.}\ }\textbf {\bibinfo {volume} {207}},\
  \bibinfo {pages} {432} (\bibinfo {year} {2016})},\ \Eprint
  {http://arxiv.org/abs/1601.01167} {arXiv:1601.01167 [hep-ph]} \BibitemShut
  {NoStop}%
\bibitem [{\citenamefont {Mertig}\ \emph {et~al.}(1991)\citenamefont {Mertig},
  \citenamefont {Bohm},\ and\ \citenamefont {Denner}}]{feynCalc3}%
  \BibitemOpen
  \bibfield  {author} {\bibinfo {author} {\bibfnamefont {R.}~\bibnamefont
  {Mertig}}, \bibinfo {author} {\bibfnamefont {M.}~\bibnamefont {Bohm}}, \ and\
  \bibinfo {author} {\bibfnamefont {A.}~\bibnamefont {Denner}},\ }\href
  {\doibase 10.1016/0010-4655(91)90130-D} {\bibfield  {journal} {\bibinfo
  {journal} {Comput. Phys. Commun.}\ }\textbf {\bibinfo {volume} {64}},\
  \bibinfo {pages} {345} (\bibinfo {year} {1991})}\BibitemShut {NoStop}%
\bibitem [{\citenamefont {Patel}(2015)}]{Patel}%
  \BibitemOpen
  \bibfield  {author} {\bibinfo {author} {\bibfnamefont {H.~H.}\ \bibnamefont
  {Patel}},\ }\href {\doibase 10.1016/j.cpc.2015.08.017} {\bibfield  {journal}
  {\bibinfo  {journal} {Comput. Phys. Commun.}\ }\textbf {\bibinfo {volume}
  {197}},\ \bibinfo {pages} {276} (\bibinfo {year} {2015})},\ \Eprint
  {http://arxiv.org/abs/1503.01469} {arXiv:1503.01469 [hep-ph]} \BibitemShut
  {NoStop}%
\end{thebibliography}%

\onecolumngrid
%\begin{widetext}

\clearpage
\section{Supplementary Material} 
%\label{sec:suppl}

\begin{center}
\emph{``Breakdown of chiral perturbation theory for the axion hot dark matter bound''} 

by Luca Di Luzio, Guido Martinelli, and Gioacchino Piazza
\end{center}

\begin{figure}[h!]
\centering
\includegraphics[width=8cm]{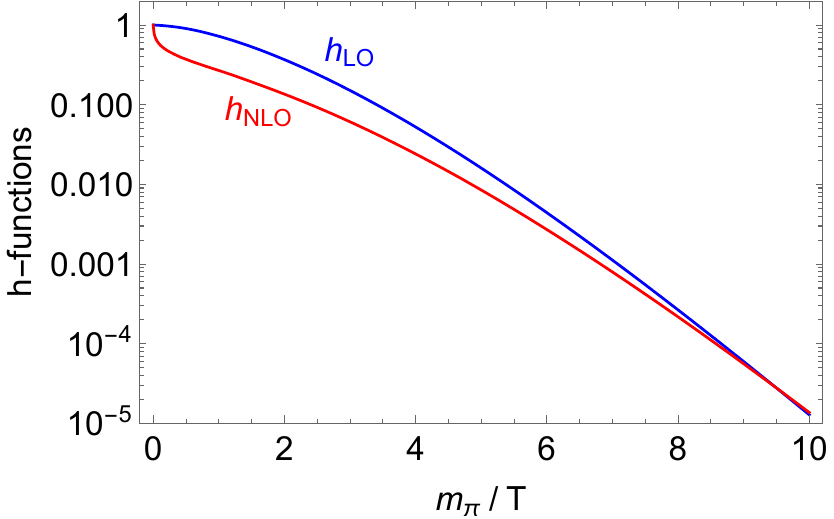}
%\vspace*{-4ex}%
\caption{Numerical profile of the $h_{\rm LO}$ and $h_{\rm NLO}$ functions.}
\label{fig:hfuncts}       
\end{figure}

The calculation of the amplitudes was carried out using the  
computational tools \textsc{FeynRules} \cite{FeynRules1,feynRules2}, \textsc{FeynArts} \cite{feynArts}, \textsc{FeynCalc} \cite{feynCalc1,feynCalc2,feynCalc3} and \textsc{Package-X} \cite{Patel}.
The full analytical expression of the 
renormalized 
NLO amplitude for the $a \piz\rightarrow\pip \pim$ process  
reads 
\begin{align} 
\label{eq:Mapi0pippimNLO} 
&\mathcal{M}^{\rm NLO}_{a \piz \rightarrow \pip \pim} =
      \frac{ C_{a\pi}}{192 \pi ^2 f_{\pi }^3  f_a } 
      \Bigg\{ 15 m_{\pi }^2 (u+t)-11 u^2-8 u t-11 t^2 
      -6 \overline{\ell_1}  \left(m_{\pi }^2-s\right) \left(2 m_{\pi }^2-s\right) \nonumber \\
       & - 6 \overline{\ell_2} \left(-3 m_{\pi }^2 (u+t)+4 m_{\pi }^4+u^2+t^2\right) 
     %  + 9 m_{\pi }^4  \overline{\ell_3 }
       +18  \overline{\ell_4} m_{\pi }^2 (m_{\pi }^2-s)
      % + 576 \pi^2 \ell_7 m_{\pi }^4 \(\frac{m_d-m_u}{m_d+m_u}\)^2
       \nonumber \\
       &+3 \[ 3  \sqrt{1 -\frac{4 m_{\pi }^2}{s}} s \left(m_{\pi }^2-s\right) \ln{\left(\frac{ \sqrt{s \left(s-4 m_{\pi }^2\right)}+2 m_{\pi }^2-s}{2 m_{\pi }^2}\right)} \right. \nonumber \\
       &+\sqrt{1- \frac{4 m_{\pi }^2}{t}} \left(m_{\pi }^2 (t-4 u)+3 m_{\pi }^4+t (u-t)\right)
       \ln{\left(\frac{\sqrt{t \left(t-4 m_{\pi }^2\right)}+2 m_{\pi }^2-t}{2 m_{\pi }^2}\right)}\nonumber \\
       &\left. +  \sqrt{ 1-\frac{4 m_{\pi }^2}{u}} \left(m_{\pi }^2 (u-4 t)+3 m_{\pi }^4+u (t-u)\right) 
       \ln{\left(\frac{ \sqrt{u \left(u-4 m_{\pi }^2\right)}+2 m_{\pi }^2-u}{2 m_{\pi }^2}\right)}\] \Bigg\} \nonumber \\
       & +\frac{4  \ell_7 m_{\pi }^2 m_d \left(s-2 m_{\pi }^2 \right) m_u \left(m_d-m_u\right)}{ f_{\pi }^3 f_a \left(m_d+m_u\right){}^3} 
  \, ,
\end{align} 
where the term proportional to $\overline{\ell_4}$ in the second row 
arises from the NLO correction to $f_\pi$ (see Ref.~\cite{Gasser:1983yg}) in the LO amplitude.
% LO amplitude, via the NLO corrections to $m_\pi$ and $f_\pi$ (see Ref.~\cite{Gasser:1983yg}).
%The latter include the charged-neutral pion mass difference arising at second order in the isospin breaking parameter $m_d-m_u$, which has been neglected in the numerical integration of the rate. 
The amplitudes for the crossed channels 
$a \pim\rightarrow\piz \pim$ and $a \pip\rightarrow\pip \piz$ 
are obtained by cross symmetry through the replacements 
$s \leftrightarrow t$ and $s \leftrightarrow u$, respectively. 
Next, we describe our procedure to analytically reduce the 12-dimensional phase space 
integral of 
%Eq.~(14) of the Letter
\eq{Gamma1} 
down to a 5-dimensional one.
We first integrate the fourth-particle phase space 
in 
%Eq.~(14) of the Letter
\eq{Gamma1} 
using the relation
$d^3 \mathbf{p}_4 / (2 E_4)= d^4 p_4 \delta\( p_4^2-m_4^2\)\theta(p_4^0)$.
Therefore, defining the angles 
$\alpha$ and $\theta$ via 
$\cos \alpha = \frac{\mathbf{p}_1 \cdot \mathbf{p}_2}{|\mathbf{p}_1| | \mathbf{p}_2|}$ 
and 
$\cos \theta = \frac{\mathbf{p}_1 \cdot \mathbf{p}_3}{|\mathbf{p}_1| | \mathbf{p}_3|}$,
the thermalization rate becomes 
\begin{align}\label{GammaInt}
    \Gamma_a &= \frac{1}{n_a^{\rm eq}} \int \frac{d p_1 |\mathbf{p}_1|^2 }{2 E_1}\frac{d p_2 |\mathbf{p}_2|^2 }{2 E_2}\frac{d p_3 |\mathbf{p}_3|^2 }{2 E_3} \int_{-1}^{1} d \cos \alpha \int_{-1}^{1} d \cos \theta \int_{0}^{2\pi} d\beta \sum|\mathcal{M}|^2 \frac{4 \pi}{(2\pi)^{7}}  \nonumber  \\
     & \times \frac{\delta \left(E_1 - \xi(E_2,E_3, \alpha, \beta , \theta) \right) }{2 |E_2-E_3-|\mathbf{p}_2| \cos \alpha + |\mathbf{p}_3| \cos \theta |}f_1 f_2 (1+ f_3)(1+ f_4) \, ,
\end{align}
with 
$\beta$ the angle between the scattering planes 
defined by $(\mathbf{p}_1,\mathbf{p}_2)$ and $(\mathbf{p}_3,\mathbf{p}_4)$, 
and the function $\xi$ given by
\beq
 \xi(E_2,  E_3, \alpha, \beta, \theta ) = \frac{2 \left(E_2 E_3-|\mathbf{p}_2| |\mathbf{p}_3| \left(\sin \alpha \sin \theta \cos  \beta  +\cos  \alpha  \cos  \theta \right)\right)-m_\pi^2}{2 (E_2-E_3-|\mathbf{p}_2| \cos  \alpha +|\mathbf{p}_3| \cos  \theta )} \, . 
 \eeq
\eq{GammaInt} is then integrated numerically, 
leading to 
\eq{gammaInterf}. 
%Eq.~(15) of the Letter.
Finally, the numerical profile of the $h$-functions entering the axion-pion thermalization rate in 
%Eq.~(15) of the Letter 
\eq{gammaInterf}
is plotted in \fig{fig:hfuncts}.

\end{document}